\documentclass[aps,prc,twocolumn,superscriptaddress,showpacs]{revtex4-1}

\usepackage{color,graphicx,amsmath,amssymb,bm}
\usepackage[T1]{fontenc}

\newcommand{\bra}[1]{\left\langle #1\right|}
\newcommand{\ket}[1]{\left| #1\right\rangle}
\newcommand{\kev}{\,\mathrm{keV}}
\newcommand{\mev}{\,\mathrm{MeV}}
\newcommand{\gev}{\,\mathrm{GeV}}
\newcommand{\gevi}{\,\mathrm{GeV}^{-1}}
\newcommand{\fm}{\,\mathrm{fm}}
\newcommand{\fmi}{\,\mathrm{fm}^{-1}}
\newcommand{\fmiq}{\,\mathrm{fm}^{-3}}

\newcommand{\km}{\,\mathrm{km}}
\newcommand{\tr}{\mathrm{Tr}}

\begin{document}

\title{Neutron matter from chiral effective field theory interactions}

\author{T.\ Kr\"uger}
\email[E-mail:~]{tkrueger@theorie.ikp.physik.tu-darmstadt.de}
\affiliation{Institut f\"ur Kernphysik, Technische Universit\"at Darmstadt, 
64289 Darmstadt, Germany}
\affiliation{ExtreMe Matter Institute EMMI, GSI Helmholtzzentrum f\"ur 
Schwerionenforschung GmbH, 64291 Darmstadt, Germany}
\author{I.\ Tews}
\email[E-mail:~]{tews@theorie.ikp.physik.tu-darmstadt.de}
\affiliation{Institut f\"ur Kernphysik, Technische Universit\"at Darmstadt,
64289 Darmstadt, Germany}
\affiliation{ExtreMe Matter Institute EMMI, GSI Helmholtzzentrum f\"ur 
Schwerionenforschung GmbH, 64291 Darmstadt, Germany}
\author{K.\ Hebeler}
\email[E-mail:~]{hebeler.4@osu.edu}
\affiliation{Department of Physics, The Ohio State University, 
Columbus, OH 43210, USA}
\author{A.\ Schwenk}
\email[E-mail:~]{schwenk@physik.tu-darmstadt.de}
\affiliation{ExtreMe Matter Institute EMMI, GSI Helmholtzzentrum f\"ur
Schwerionenforschung GmbH, 64291 Darmstadt, Germany}
\affiliation{Institut f\"ur Kernphysik, Technische Universit\"at Darmstadt,
64289 Darmstadt, Germany}

\begin{abstract}
The neutron-matter equation of state constrains the properties of many
physical systems over a wide density range and can be studied
systematically using chiral effective field theory (EFT). In chiral
EFT, all many-body forces among neutrons are predicted to
next-to-next-to-next-to-leading order (N$^3$LO). We present details
and additional results of the first complete N$^3$LO calculation of
the neutron-matter energy, which includes the subleading three-nucleon
as well as the leading four-nucleon forces, and provides theoretical
uncertainties. In addition, we discuss the impact of our results for
astrophysics: for the supernova equation of state, the symmetry energy
and its density derivative, and for the structure of neutron
stars. Finally, we give a first estimate for the size of the N$^3$LO
many-body contributions to the energy of symmetric nuclear matter,
which shows that their inclusion will be important in nuclear
structure calculations.
\end{abstract}


\maketitle

\section{Introduction}

Chiral effective field theory (EFT) provides a systematic expansion
for nuclear forces including theoretical uncertainties~\cite{RMP},
where the development and applications of three-nucleon (3N) forces
are a frontier~\cite{3NRMP}. In this context, neutron matter
constitutes a unique laboratory for chiral EFT, because all many-body
forces are predicted to N$^3$LO~\cite{fullN3LO}. This offers the
possibility to provide reliable constraints based on chiral EFT
interactions for neutron-rich matter in astrophysics, for the equation
of state, the symmetry energy and its density dependence, and for the
structure of neutron stars~\cite{nm,nstar}, but also allows us to test the
chiral EFT power counting and the hierarchy of many-body forces over a
wide density range. In addition, the prediction of many-body forces
makes neutron-rich nuclei very exciting to test chiral EFT
interactions against experiments at rare isotope beam
facilities~\cite{Oxygen,Holt1,Gallant,Hagen1,Lunderberg,Hagen2,Roth,%
Petri,Holt2,protonrich,R3B,Hergert,26F,IMSRGHeiko2,Carlo,pairing}.

Neutron matter has been studied in chiral EFT using lattice
simulations~\cite{NLOlattice} and based on in-medium chiral
perturbation theory~\cite{Wolfram1,Wolfram2}. In addition, neutron
matter has been calculated using renormalization-group-evolved chiral
EFT interactions~\cite{nm}, where the renormalization group (RG)
evolution improves the convergence of the many-body expansion around
the Hartree-Fock energy~\cite{nucmatt1,PPNP} and in a chiral Fermi
liquid approach~\cite{Jeremy}. These studies demonstrated that 3N
forces are significant at nuclear densities and that the dominant
uncertainty is due to the truncation of 3N forces at the
next-to-next-to-leading-order (N$^2$LO) level~\cite{nm}. Moreover,
first Quantum Monte Carlo calculations with chiral EFT interactions
are providing nonperturbative benchmarks for neutron matter at nuclear
densities~\cite{QMC_chiral}.

Motivated by these studies and by the derivation of the parameter-free
N$^3$LO 3N and four-nucleon (4N)
interactions~\cite{IR,N3LOlong,N3LOshort,4N,4Nlong}, we recently
presented the first calculation of the neutron-matter energy that
includes all two-nucleon (NN), 3N and 4N forces consistently to
N$^3$LO~\cite{fullN3LO}. In this paper, we discuss details of our
complete N$^3$LO calculation and present additional results as well as
applications to astrophysics, for the equation of state and for the
mass-radius relation of neutron stars. In addition, we give a first
estimate for the size of the N$^3$LO many-body contributions to the
energy of symmetric nuclear matter in the Hartree-Fock
approximation. This presents only a first step towards a complete
calculation of nuclear matter, where contributions from many-body
forces beyond Hartree Fock are considerably more important than for
neutron matter~\cite{nucmatt2}. Our first results show that the
inclusion of N$^3$LO 3N forces will be important in nuclear structure
calculations.

This paper is organized as follows. In Sec.~\ref{sec:interaction}
we discuss the chiral EFT interactions included in this work. Details
of the many-body calculation and convergence are given in
Sec.~\ref{sec:mbdetails}. Our results for neutron matter are
presented in Sec.~\ref{sec:results}, including a detailed discussion
of the uncertainties. In Sec.~\ref{sec:applications}, we apply our
results to the equation of state, in particular to the symmetry energy
and its density dependence, and discuss the resulting constraints for
the structure of neutron stars. We show first results for the N$^3$LO
3N and 4N contributions in symmetric nuclear matter at the
Hartree-Fock level in Sec.~\ref{sec:symmetric}. Finally, we
summarize and give an outlook.

\section{Chiral EFT interactions}
\label{sec:interaction}

\subsection{N$^\text{2}$LO and N$^\text{3}$LO NN forces}
\label{sec:potentials}

The largest interaction contributions to the neutron-matter energy
arise from NN forces. For our past applications of chiral EFT
interactions to nucleonic matter~\cite{nm,nucmatt2}, the RG evolution
has been used to evolve NN potentials to low-momentum interactions to
improve the many-body convergence~\cite{nucmatt1,PPNP}. In this work,
we present calculations based directly on chiral EFT interactions
without RG evolution and study the perturbative convergence following
Ref.~\cite{fullN3LO}.

We investigate all existing NN potentials at N$^2$LO and at N$^3$LO of
Epelbaum, Gl\"ockle, and Mei{\ss}ner (EGM)~\cite{EGM,EGM2} with
cutoffs $\Lambda/\widetilde{\Lambda} = 450/500$, $450/700$, $550/600$,
$600/600$ and $600/700\mev$, where $\Lambda$ and $\widetilde{\Lambda}$
denote the cutoff in the Lippmann-Schwinger equation and in the
two-pion-exchange spectral-function regularization, respectively; as
well as the available N$^3$LO NN potentials of Entem and Machleidt
(EM)~\cite{EM,EMRept} with cutoffs $\Lambda = 500$ and $600\mev$. The
EM 500 MeV potential is most commonly used in nuclear structure
calculations, while the EGM potentials have only been studied in some
many-body calculations~\cite{nucmatt2}, although they allow to explore
a wider cutoff range.

The N$^3$LO 3N and 4N forces involve the momentum-independent NN
contact interactions $C_S + C_T \bm{\sigma}_1 \cdot \bm{\sigma}_2$. In
particular, they mainly depend on $C_T$. The $C_S$ and $C_T$ values of
the different N$^3$LO NN potentials are listed in Table~\ref{tab:csct}
for the neutron-proton case (the charge dependence contributes to
higher-order charge-dependent 3N forces). For a perturbative
calculation, we require Wigner symmetry ($C_T = 0$) to be fulfilled
approximately at the interaction level. This is not the case for the
EGM potentials with cutoffs 600/600 and 600/700 MeV, which have large
spin-dependent couplings $C_T \sim C_S$ (and even a repulsive
spin-independent $C_S$), and would lead to large $C_T$-dependent 3N
forces at N$^3$LO.

\begin{table}[t]
\caption{Spin-independent and spin-dependent two-body contact 
couplings $C_S$ and $C_T$, respectively, for the N$^3$LO NN 
potentials of Refs.~\cite{EGM2,EM,EMRept}.\label{tab:csct}}
\begin{tabular*}{\columnwidth}{@{\extracolsep{\fill}}lcc}
\hline\hline
NN potential & $C_S\,[\!\fm^{2}]$ & $C_T\,[\!\fm^{2}]$\\
\hline
EGM 450/500 MeV \cite{EGM2} & $-4.19$ & $-0.45$\\
EGM 450/700 MeV \cite{EGM2} & $-4.71$ & $-0.24$\\
EM 500 MeV \cite{EM,EMRept} & $-3.90$ & $0.22$\\
\hline
EGM 550/600 MeV \cite{EGM2} & $-1.24$ & $0.36$\\
EGM 600/600 MeV \cite{EGM2} & $3.45$ & $2.07$\\
EGM 600/700 MeV \cite{EGM2} & $1.31$ & $1.00$\\
EM 600 MeV \cite{EM,EMRept} & $-3.88$ & $0.28$\\
\hline\hline
\end{tabular*}
\end{table}

\subsection{N$^\text{2}$LO 3N forces}

Three-nucleon forces enter at N$^2$LO in the chiral EFT expansion
without explicit Deltas~\cite{Kolck,N2LO3N}. Due to the Pauli principle
and the coupling of pions to spin, only the $c_1$ and $c_3$ parts of
the long-range two-pion-exchange 3N interactions contribute at
N$^2$LO~\cite{nm} (see also Ref.~\cite{Tolos}). The same $c_i$
couplings also enter NN interactions at N$^2$LO and have been
determined from pion-nucleon or NN scattering. The $c_1$ and $c_3$ values
used in chiral NN potentials are given in Table~\ref{tab:c_i}.
Note, however, that the range adopted in the NN potentials of
Table~\ref{tab:c_i} does not reflect the allowed range for the $c_i$
couplings, which are not satisfactorily constrained at present, e.g.,
with a range of $c_3 = -(3.2\,$--$\,5.9) \gevi$ from different
theoretical analyses (see Table~I in Ref.~\cite{3NRMP}).

\begin{table}[t]
\caption{Values of couplings $c_1$ and $c_3$ for the different NN potentials,
as well as from Krebs, Gasparyan, and Epelbaum (KGE, Ref.~\cite{Krebs}),
and the range adopted in this work.\label{tab:c_i}}
\begin{tabular*}{\columnwidth}{@{\extracolsep{\fill}}lcc}
\hline\hline
& $c_1\,[\!\gev^{-1}]$ & $c_3\,[\!\gev^{-1}]$ \\
\hline
N$^2$LO/N$^3$LO EGM NN \cite{EGM,EGM2} & $-0.81$ & $-3.40$ \\
N$^3$LO EM NN \cite{EM,EMRept} & $-0.81$ & $-3.20$ \\
\hline
N$^2$LO KGE \cite{Krebs} & $-(0.26\,$--$\,0.58)$ & $-(2.80\,$--$\,3.14)$ \\
`N$^2$LO' KGE (recom.) \cite{Krebs} & $-(0.37\,$--$\,0.73)$ & $-(2.71\,$--$\,3.38)$ \\
N$^3$LO KGE \cite{Krebs} & $-(0.75\,$--$\,1.13)$ & $-(4.77\,$--$\,5.51)$ \\
\hline
N$^2$LO this work & $-(0.37\,$--$\,0.81)$ & $-(2.71\,$--$\,3.40)$ \\
N$^3$LO this work & $-(0.75\,$--$\,1.13)$ & $-(4.77\,$--$\,5.51)$ \\
\hline\hline
\end{tabular*}
\end{table}

We see from Table~\ref{tab:c_i} that, while the $c_1$ value is of
natural size, the $c_3$ value is large. This is due to the
single-$\Delta$ excitation, which enhances $c_3 \sim 1/(m_\Delta-m)$ by
the $\Delta$-nucleon mass difference to a large value ($c_1=0$ for a
single-$\Delta$ excitation). In chiral EFT with explicit $\Delta$'s, the
single-$\Delta$ contribution would in fact be included at one order
lower; at next-to-leading order (NLO) in this case. The large $c_3$
value has two effects. First, it leads to a slower convergence at the
order when the $c_i$ contributions enter. This corresponds to
topologies where $\Delta$ excitations are important. This can already be
seen in the convergence pattern with NN interactions, where the
leading two-pion-exchange NN interaction at NLO receives large
contributions due to the large $c_i$ that enter the subleading
two-pion-exchange NN interaction at
N$^2$LO~\cite{RMP,PPNPEpelbaum}. Therefore, for 3N and 4N forces
important contributions to the N$^3$LO interactions studied here can
be expected in topologies where the $c_i$ couplings enter at
N$^4$LO~\cite{Krebs,Kaiser}. This convergence pattern can be improved
by including the $\Delta$ explicitly in chiral EFT. Second, the large
$c_3$ coupling in the N$^2$LO 3N interaction also worsens the
perturbative convergence of the many-body expansion around the
Hartree-Fock energy. This is most important for the large $c_3$ values
considered in the N$^3$LO calculation of this work (see Table~\ref{tab:c_i}).

In addition, we list in Table~\ref{tab:c_i} the $c_i$ values extracted
from a high-order analysis up to N$^4$LO of Krebs, Gasparyan, and
Epelbaum (KGE, Ref.~\cite{Krebs}). The KGE ranges at N$^2$LO and
N$^3$LO are given in Table~\ref{tab:c_i}, in addition to values
recommended to be used in an N$^2$LO calculation that are tuned to
capture the higher-order result. In this work, we take the KGE
recommended $c_i$ range for the N$^2$LO calculation, minimally
enlarged to include the $c_i$ values of the NN potentials, and the KGE
N$^3$LO $c_i$ range for our complete N$^3$LO calculation. Note the
large $c_3$ value for the latter, which is still in the range of
Table~I in Ref.~\cite{3NRMP}. We thus explore $c_i$ values in the
many-body interactions without varying the $c_i$ in the NN
potential. This is because changing the $c_i$ in the NN potential
would also require an adjustment of other couplings in the fit to NN
data. We expect that some of the changes can be absorbed by the
N$^3$LO NN contact interactions, but it is very important to develop
new N$^3$LO NN potentials that can explore this sensitivity.

\subsection{N$^3$LO 3N and 4N forces}

The many-body forces at N$^3$LO are predicted by couplings in previous
orders of the chiral EFT expansion. Hence, there are no new parameters
for N$^3$LO 3N and 4N interactions~\cite{RMP}.  The subleading N$^3$LO
3N forces have been derived recently~\cite{IR,N3LOlong,N3LOshort}.
They can be grouped into five topologies, where the latter two depend
on the NN contact couplings $C_T$ and $C_S$ (see the Appendix):
\begin{equation}
V_{\text{3N}}^{\text{N}^3\text{LO}} = V^{2\pi} + V^{2\pi\text{-}1\pi} + 
V^{\text{ring}} + V^{2\pi\text{-cont}} + V^{1/m} \,.
\end{equation}
Here, $V^{2\pi}$, $V^{2\pi\text{-}1\pi}$, and $V^{\text{ring}}$ denote
the long-range two-pion-exchange, the two-pion--one-pion-exchange, and
the pion-ring 3N interactions, respectively~\cite{N3LOlong}. The terms
$V^{2\pi\text{-cont}}$ and $V^{1/m}$ are the short-range
two-pion-exchange--contact 3N interaction and 3N relativistic
corrections, respectively~\cite{N3LOshort}. The latter are
small~\cite{fullN3LO} and depend also on the constants $\bar{\beta}_8$
and $\bar{\beta}_9$, which need to be chosen consistently with the
unitary transformation used for the NN potentials~\cite{N3LOshort}. In
addition, there could be short-range one-pion-exchange--contact 3N
interactions, but they have been shown to vanish at
N$^3$LO~\cite{N3LOshort}.

According to the chiral power counting, 4N forces enter at
N$^3$LO. They have been derived in Refs.~\cite{4N,4Nlong} and depend
also on the contact coupling $C_T$, but in neutron matter the
$C_T$-dependent parts do not contribute. There are seven 4N topologies
that lead to non-vanishing contributions. In neutron matter
only two three-pion-exchange diagrams (in Ref.~\cite{4N} named $V^a$
and $V^e$) and the pion-pion-interaction diagram ($V^f$)
contribute~\cite{fullN3LO}.

\section{Many-body details}
\label{sec:mbdetails}

\subsection{Hartree Fock}
\label{sec:mbdetailsHF}

\begin{figure}[t]
\centering
\includegraphics[width=0.9\columnwidth]{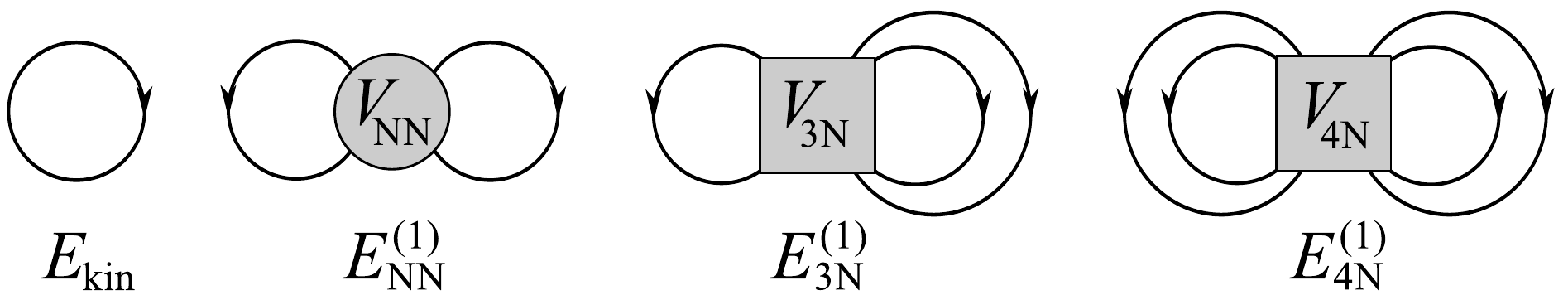}
\caption{Diagrams contributing to the Hartree-Fock energy. These
include the kinetic energy $E_\text{kin}$ and the first-order NN, 
3N, and 4N interaction energies $E_\text{NN}^{(1)}$, $E_\text{3N}^{(1)}$,
and $E_\text{4N}^{(1)}$.\label{fig:diagrams_hf}}
\end{figure}

We calculate the energy per particle at the Hartree-Fock level and
include contributions beyond Hartree Fock using many-body perturbation
theory~\cite{nm,nucmatt1,nucmatt2}. The Hamiltonian is given by $H = T
+ V_\text{NN} + V_\text{3N} + V_\text{4N}$, where $T$ is the kinetic
energy and $V_\text{NN}$, $V_\text{3N}$, and $V_\text{4N}$ denote the
NN, 3N, and 4N interactions, respectively. The Hartree-Fock
contributions are shown diagrammatically in Fig.~\ref{fig:diagrams_hf}.
At this level, the contribution of the $A$-nucleon interaction to the
energy per particle is given by
\begin{multline}
\frac{E^{(1)}_{A\text{N}}}{N} = \frac{1}{n} \frac{1}{A!}\!\sum_{\sigma_1,\ldots,\sigma_A}\!\int\!\frac{d{\bf k}_1}{(2\pi)^3} \cdots \!\int\!\frac{d{\bf k}_A}{(2\pi)^3}\, f^2_R \, n_{{\bf k}_1} \cdots \, n_{{\bf k}_A} \\
\times \bra{1\ldots A} \mathcal{A}_{A} \! \sum\limits_{i_1 \neq \ldots \neq i_A}^A \! V_{A\text{N}}(i_1,\ldots,i_A) \ket{1\ldots A}\,,
\label{eq:hf}
\end{multline}
with density $n$ and short-hand notation $i \equiv {\bf
k}_i\sigma_i$. Here, $\mathcal{A}_A$ denotes the $A$-body
antisymmetrizer and $n_{{\bf k}_i}=\theta(k_{\mathrm{F}} -k_i)$ the
Fermi-Dirac distribution at zero temperature. For the many-body
forces, we use a Jacobi-momenta regulator. In terms of ${\bf k}_i$,
this is given by
\begin{equation}
 f_R = e^{-[(k_1^2+\ldots+k_A^2-{\mathbf{k}_1\cdot\mathbf{k}_2} - \ldots - \mathbf{k}_{A-1} \cdot \mathbf{k}_A)/(A \Lambda^2)]^{n_\mathrm{exp}}}\,,
\label{eq:regulator}
\end{equation}
where we take $n_\textrm{exp} =4$ and consider 3N/4N cutoffs
$\Lambda=2-2.5 \fmi$. This cutoff range allows to probe the
sensitivity to short-range many-body forces within the limits of the
employed power counting. For the evaluation of 3N/4N forces, we use
for the nucleon and pion mass, $m=938.92\mev$ and
$m_{\pi}=138.04\mev$, for the axial coupling $g_A=1.29$, and for the
pion decay constant $f_\pi=92.4\mev$~\cite{N2LO3N,N3LOlong,N3LOshort,4N,4Nlong}.

As an example, we present details of the derivation of the
Hartree-Fock energy from the N$^3$LO two-pion-exchange 3N
interactions. Their contributions can be grouped into two parts: one that
shifts the $c_i$ couplings of the N$^2$LO 3N forces and a part
\begin{align}
V_{2\pi}^{(4)} &= \frac{g_A^4}{256\pi f_{\pi}^6} \sum_{i \neq j \neq k} \frac{(\bm{\sigma}_i\cdot \mathbf{q}_i) (\bm{\sigma}_j\cdot \mathbf{q}_j)}{(\mathbf{q}_i^2+m_{\pi}^2)(\mathbf{q}_j^2+m_{\pi}^2)}\nonumber\\
&\quad \times \Bigl[ m_{\pi}(m_{\pi}^2+3q_i^2+3q_j^2+4\mathbf{q}_i\cdot \mathbf{q}_j) \nonumber\\
&\quad + (2m_{\pi}^2+q_i^2+ q_j^2+2\mathbf{q}_i\cdot \mathbf{q}_j)\nonumber\\
&\quad \times (3m_{\pi}^2+3q_i^2+3q_j^2+4\mathbf{q}_i \cdot \mathbf{q}_j)A(q_k)\Bigr] \,, \nonumber \\[1mm]
&= \sum_{i \neq j \neq k} (\bm{\sigma}_i\cdot \mathbf{q}_i) (\bm{\sigma}_j \cdot \mathbf{q}_j) F_{2\pi}^{(4)}(\mathbf{q}_i,\mathbf{q}_j) \,.
\end{align}
For the isospin part we have used that for neutrons
\begin{align}
\bra{nnn}\bm{\tau}_i \cdot \bm{\tau}_j \ket{nnn} &= 1 \,,\\[1mm]
\bra{nnn}\bm{\tau}_i \cdot \bm{\tau}_j \times \bm{\tau}_k\ket{nnn} &= 0 \,,
\end{align}
and introduced the function
$F_{2\pi}^{(4)}(\mathbf{q}_i,\mathbf{q}_j)$, which absorbs all parts
of the interaction except for the spin dependencies. Furthermore, ${\bf
q}_i = {\bf k}_i' - {\bf k}_i$ and for $F_{2\pi}^{(4)}$ we use
$\mathbf{q}_1+\mathbf{q}_3=-\mathbf{q}_2$ due to momentum
conservation. Since the particles $i$, $j$, $k$ are all neutrons and
we sum over all possible spin states, the six different terms in the
sum lead to identical contributions and we can write
\begin{equation}
V_{2\pi}^{(4)} = 6\,(\bm{\sigma}_1\cdot \mathbf{q}_1) (\bm{\sigma}_3\cdot \mathbf{q}_3)\, F_{2\pi}^{(4)}(\mathbf{q}_1,\mathbf{q}_3) \,.
\end{equation}

For the spin trace
$\tr_\sigma\bra{123}\mathcal{A}_3V_{2\pi}^{(4)}\ket{123}$ we use that
Pauli matrices are traceless and the relation $\sigma_i^a \sigma_i^b
=\delta^{ab}+ i \epsilon^{abc} \sigma_i^c$. Thus, only the parts of
the antisymmetrizer that contain the same-particle Pauli matrices as
the potential need to be considered. In this case, the terms must
contain $\bm{\sigma}_1$ and $\bm{\sigma}_3$ but not
$\bm{\sigma}_2$. The antisymmetrizer is given by
\begin{align}
\mathcal{A}_3 = 1 - P_{12} - P_{13} - P_{23} + P_{12}P_{23} + P_{13}P_{23}\,,
\end{align}
with $P_{ij} = P^k_{ij} \,
\frac{1+\bm{\sigma}_i\cdot\bm{\sigma}_j}{2}$, where $P^k_{ij}$
exchanges the momenta of particles $i$ and $j$. The last two terms can
be written as
\begin{align}
P_{12}P_{23} = \frac{1}{4}P^k_{12} P^k_{23} (1&+\bm{\sigma}_1\cdot \bm{\sigma}_2 + \bm{\sigma}_2\cdot\bm{\sigma}_3\nonumber\\
& + \bm{\sigma}_1\cdot\bm{\sigma}_3+i \bm{\sigma}_1\cdot\bm{\sigma}_3 \times\bm{\sigma}_2)\,, \nonumber \\
P_{13}P_{23}\nonumber = \frac{1}{4}P^k_{13}P^k_{13} (1& + \bm{\sigma}_1\cdot\bm{\sigma}_2 + \bm{\sigma}_2\cdot\bm{\sigma}_3\nonumber\\
&+ \bm{\sigma}_1\cdot\bm{\sigma}_3 + i \bm{\sigma}_1\cdot\bm{\sigma}_2\times\bm{\sigma}_3)\,.
\end{align}
Thus, the only relevant terms of the antisymmetrizer are
\begin{equation}
\left(-\frac{P^k_{13}}{2} + \frac{P^k_{12} P^k_{23}}{4}  + \frac{P^k_{13} P^k_{23}}{4}\right)\bm{\sigma}_1\cdot\bm{\sigma}_3 \,.
\end{equation}
Multiplying this spin part with the potential leads to
\begin{align}
\tr_{\sigma} \left[\bm{\sigma}_1\cdot\bm{\sigma}_3 V_{2\pi}^{(4)}\right] &= \tr_{\sigma}\left[ 6\, \sigma_1^a \sigma_3^a\,\sigma_1^b q_1^b\,\sigma_3^c q_3^c\, F_{2\pi}^{(4)}(\mathbf{q}_1,\mathbf{q}_3)\right] \,, \nonumber \\
&= \tr_{\sigma}\Bigl[ 6(\delta^{ab}+ i \epsilon^{abd} \sigma_1^d)(\delta^{ac}+ i \epsilon^{ace} \sigma_3^e)\nonumber\\
&\quad \times q_1^b q_3^c\, F_{2\pi}^{(4)}(\mathbf{q}_1,\mathbf{q}_3)\Bigr] \,.
\end{align}
All terms containing Pauli matrices vanish when taking the trace, so
\begin{align}
\tr_{\sigma}\left[\bm{\sigma}_1\cdot\bm{\sigma}_3 V_{2\pi}^{(4)}\right] = 8\cdot6\,\mathbf{q}_1\cdot\mathbf{q}_3\, F_{2\pi}^{(4)}\left(\mathbf{q}_1,\mathbf{q}_3\right).
\end{align}
Thus, we obtain
\begin{align}
\tr_{\sigma} \mathcal{A}_3V_{2\pi}^{(4)} &= 8 \cdot 6\,\left(-\frac{P^k_{13}}{2} + \frac{P^k_{12} P^k_{23}}{4} + \frac{P^k_{13} P^k_{23}}{4} \right)\nonumber\\
&\quad \times \mathbf{q}_1\cdot\mathbf{q}_3 \, F_{2\pi}^{(4)}(\mathbf{q}_1,\mathbf{q}_3) \,.
\end{align}
Putting everything together yields for the spin-summed antisymmetrized
matrix element
\begin{widetext}
\begin{align}
\langle V_{2\pi}^{(4)}\rangle &= \frac{1}{3!} \, \tr_{\sigma}\bra{123} \mathcal{A}_{3} V_{2\pi}^{(4)} \ket{123} = 8 \bra{123}\!\left(-\frac{P^k_{13}}{2} + \frac{P^k_{12} P^k_{23}}{4} + \frac{P^k_{13} P^k_{23}}{4} \right)\! \,\mathbf{q}_1 \cdot \mathbf{q}_3 \, F_{2\pi}^{(4)}(\mathbf{q}_1,\mathbf{q}_3) \ket{123} \,, \nonumber \\
&=-4 \, \mathbf{k}_{31}\cdot \mathbf{k}_{13} \, F_{2\pi}^{(4)}(\mathbf{k}_{31},\mathbf{k}_{13}) + 2 \, \mathbf{k}_{21}\cdot \mathbf{k}_{13} \, F_{2\pi}^{(4)}(\mathbf{k}_{21},\mathbf{k}_{13}) + 2\, \mathbf{k}_{31} \cdot \mathbf{k}_{23}\, F_{2\pi}^{(4)}(\mathbf{k}_{31},\mathbf{k}_{23}) \,, \nonumber \\
&= \, 4 \left[k_{13}^2 \, F_{2\pi}^{(4)}(-\mathbf{k}_{13},\mathbf{k}_{13}) - \mathbf{k}_{12} \cdot \mathbf{k}_{13} \, F_{2\pi}^{(4)}(-\mathbf{k}_{12},\mathbf{k}_{13})\right] \,,
\end{align}
\end{widetext}
where $\mathbf{k}_{ij}=\mathbf{k}_i-\mathbf{k}_j$, and we have
relabeled the momentum indices in the last step, because the momentum
integrals are equal for the three neutrons and the regulator is
symmetric under exchange of the momenta.

Analogously, we obtain the expressions for the other N$^3$LO 3N- and
4N-interaction matrix elements at the Hartree-Fock level. They are
given in Appendix~\ref{app:nm}. The analytic derivations have been
checked independently and by using an automated Mathematica routine for
the spin traces.

\subsection{Beyond Hartree Fock}

\begin{figure}[t]
\centering
\includegraphics[width=0.95\columnwidth]{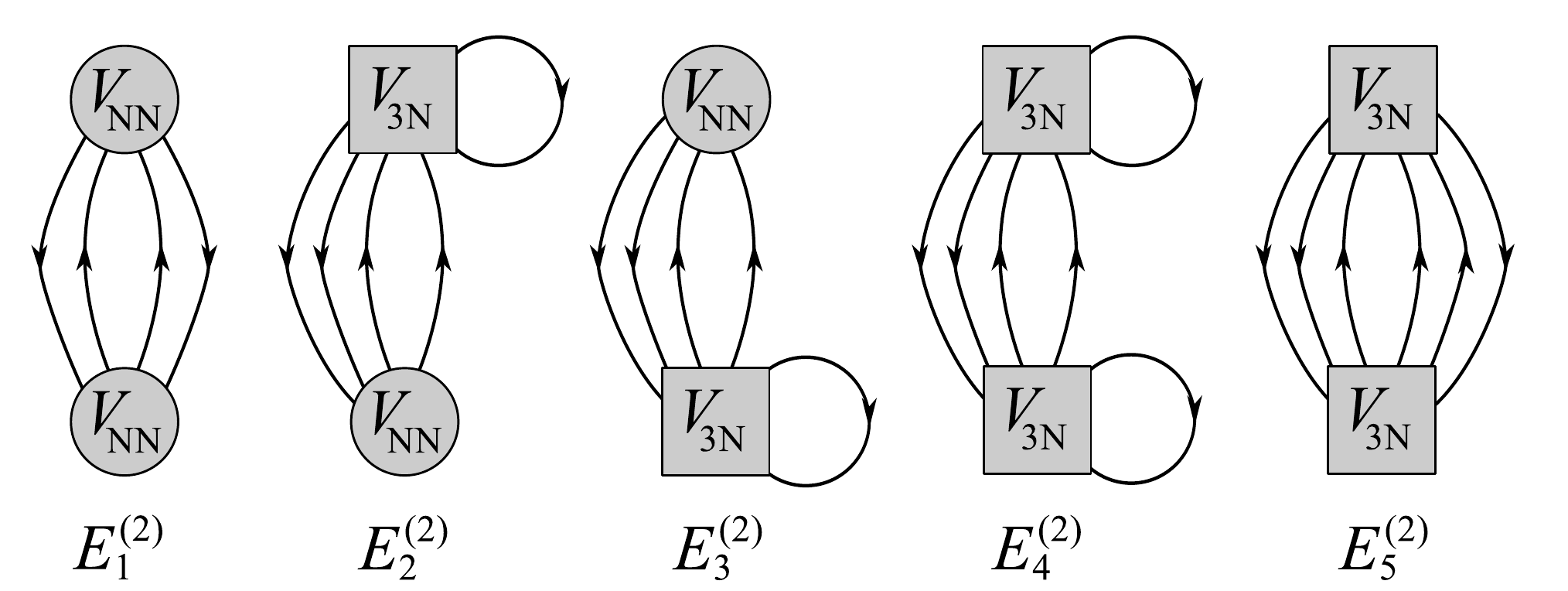}
\caption{Second-order contributions to the energy due to NN-NN 
correlations, $E_1^{(2)}$, NN-3N and 3N-3N correlations $E_{2/3}^{(2)}$
and $E_4^{(2)}$, where the 3N forces enter as density-dependent
two-body interactions, as well as the residual 3N-3N contribution
$E_5^{(2)}$ not considered here.\label{fig:diagrams_2nd}}
\end{figure}

For nucleonic matter based on chiral EFT interactions, contributions
beyond the Hartree-Fock level are
important~\cite{nm,nucmatt1,nucmatt2}. The dominant contribution to
the energy is due to NN-NN correlations [$E^{(2)}_1$]. In addition,
there are NN-3N correlations [$E^{(2)}_2$ and $E^{(2)}_3$], 3N-3N
correlations [$E^{(2)}_4$ and $E^{(2)}_5$], where the $E_i^{(2)}$
follow the notation of Fig.~\ref{fig:diagrams_2nd}, as well as
NN-4N, 3N-4N and 4N-4N correlations. Based on the results of
Refs.~\cite{fullN3LO,nm}, we expect the residual 3N-3N contribution
$E^{(2)}_5$ and all contributions including 4N interactions to be small.

The second-order contribution to the energy due to NN interactions and
including 3N interactions as density-dependent two-body interactions
is given by
\begin{align}
\sum_{i=1}^4E^{(2)}_i &= \frac{1}{4}\left[\prod_{i=1}^4 \sum_{\sigma_i}\int \frac{d^3{\bf k}_i}{(2\pi)^3}\right]\left|\bra{12}V_\text{as}^{(2)}\ket{34}\right|^2 \nonumber\\
&\quad \times \frac{n_{\mathbf{k}_1}n_{\mathbf{k}_2}(1-n_{\mathbf{k}_3})(1-n_{\mathbf{k}_4})}{\varepsilon_{\mathbf{k}_1}+\varepsilon_{\mathbf{k}_2}-\varepsilon_{\mathbf{k}_3}-\varepsilon_{\mathbf{k}_4}}\nonumber\\
&\quad \times (2\pi)^3\delta(\mathbf{k}_1+\mathbf{k}_2-\mathbf{k}_3-\mathbf{k}_4)\,,
\label{eq:2ndhf}
\end{align}
where $V_\text{as}^{(2)} = (1-P_{12})V_{\text{NN}} +
\overline{V}_\text{3N}$ is the antisymmetrized two-body interaction,
which includes NN interactions and density-dependent two-body
interactions from N$^2$LO 3N forces~\cite{nm}. The latter are obtained
by summing the third particle over the occupied states in the Fermi
sea
\begin{align}
\overline{V}_\text{3N} = \sum_{\sigma_3}\int\frac{d^3{\bf k}_3}{(2\pi)^3}\,n_{\mathbf{k}_3}\mathcal{A}_3 V_\text{3N}^{\text{N}^2\text{LO}}\Bigg|_\text{nnn} \,.
\end{align}

At third order, we include particle-particle diagrams as in
Ref.~\cite{nucmatt2}. Their size provides a test of the convergence of
the many-body calculation. We divide the third-order particle-particle
contributions into classes $E^{(3)}_i$, which are based on the
$E_i^{(2)}$ of Fig.~\ref{fig:diagrams_2nd} by adding one additional
ladder and vertex with anti-symmetrized effective two-body
interactions $V_\text{as}^{(2)} = (1-P_{12})V_{\text{NN}} +
\overline{V}_\text{3N}$ to the different diagrams $E^{(2)}_i$.

\subsection{Convergence}

\begin{figure*}[p]
\centering
\includegraphics[width=0.7\textwidth]{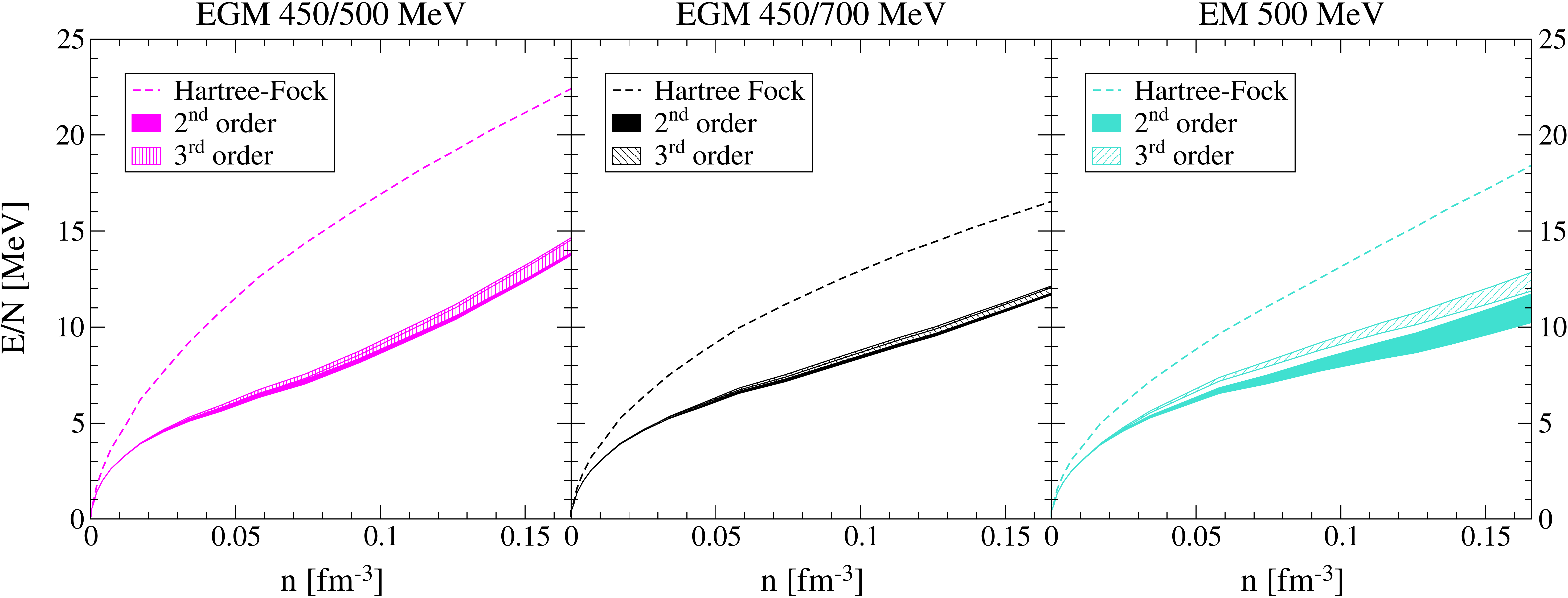}\\
\vspace{0.2cm}
\includegraphics[width=0.9\textwidth]{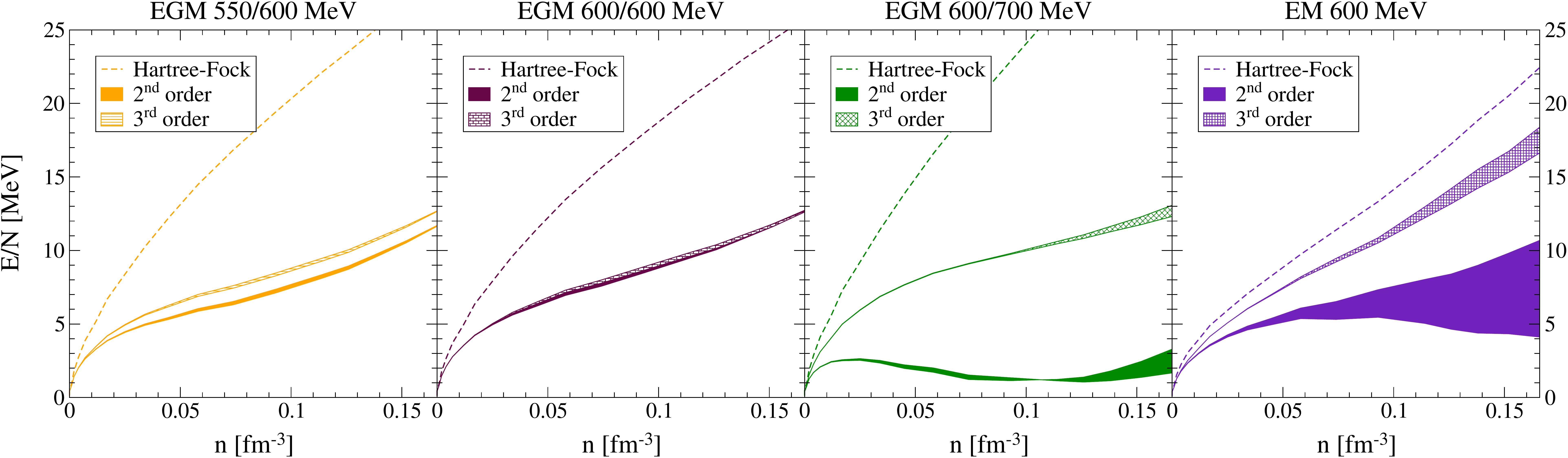}
\caption{(Color online) Energy per particle as a function of density
for the different N$^3$LO NN potentials of Refs.~\cite{EGM,EGM2,EM,EMRept}.
The dashed lines are Hartree-Fock results only. The filled and shaded bands
are second- and third-order energies, respectively, where at each order
the band ranges from using a free to a Hartree-Fock spectrum.
\label{fig:NNconvergence}}
\vspace{0.4cm}
\includegraphics[width=0.7\textwidth]{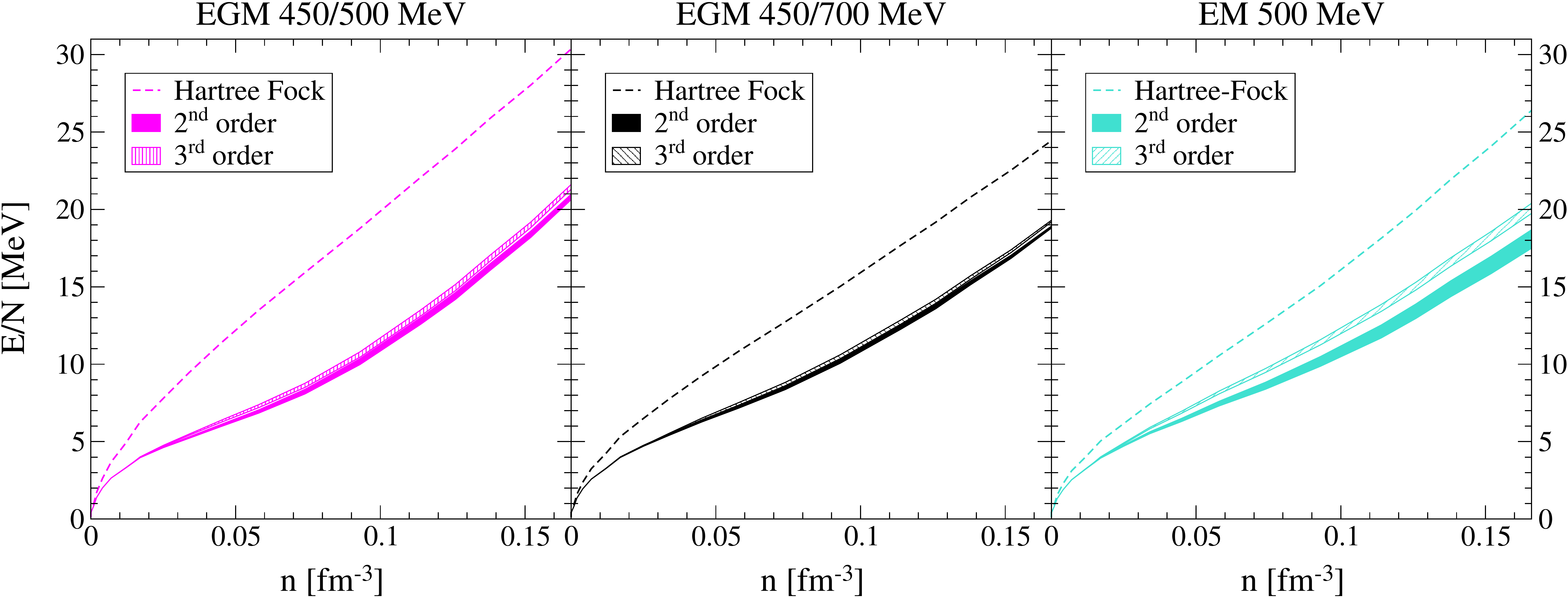}\\
\vspace{0.2cm}
\includegraphics[width=0.9\textwidth]{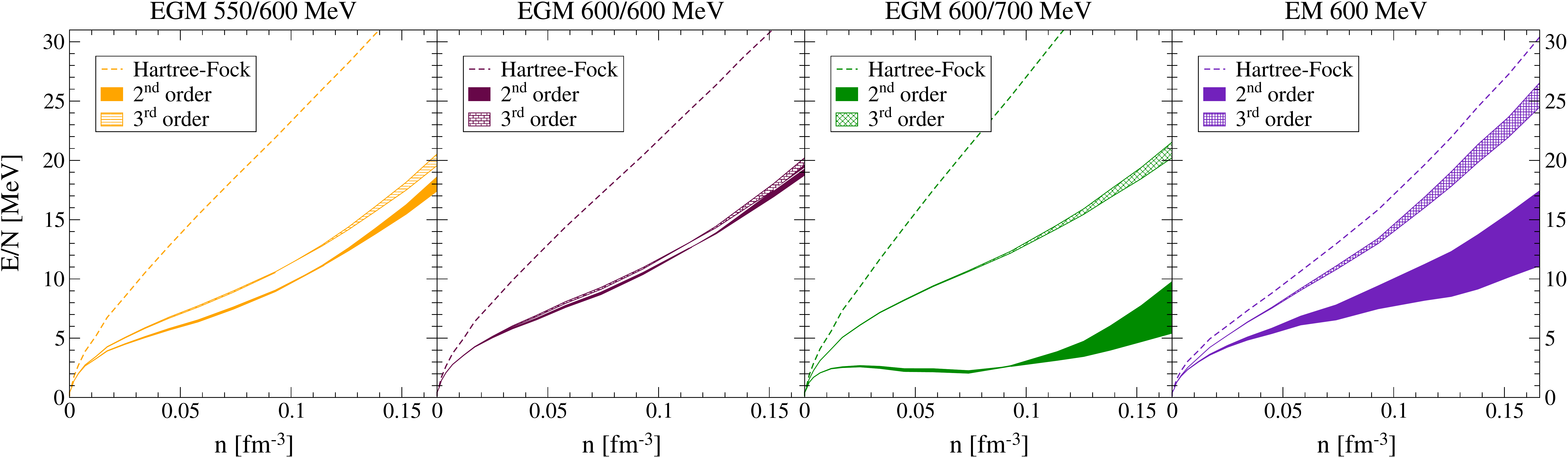}
\caption{(Color online) Energy per particle as a function of density
for the different N$^3$LO NN potentials of Refs.~\cite{EGM,EGM2,EM,EMRept}
and including the leading N$^2$LO 3N forces. The dashed lines are 
Hartree-Fock results only. The filled and shaded bands are second- and 
third-order energies, respectively, where at each order the band ranges
from using a free to a Hartree-Fock spectrum. All calculations are 
performed for a 3N cutoff $\Lambda=2.0\fmi$ and low-energy couplings
$c_1 = 0.75\gevi$ and $c_3 =4.77\gevi$.\label{fig:NN3Nconvergence}}
\end{figure*}

To study the perturbative convergence of the different NN potentials,
we calculate the Hartree-Fock as well as second- and third-order
energies with both free and Hartree-Fock single-particle energies.
First, we consider NN interactions only and then study the changes
when including also N$^2$LO 3N forces. The results are shown in
Figs.~\ref{fig:NNconvergence} and~\ref{fig:NN3Nconvergence},
respectively. The bands at each order range from using a free to a
Hartree-Fock single-particle spectrum. In addition, we give in
Table~\ref{tab:convergence} the maximal difference between the
Hartree-Fock-spectrum results at second order and those with a free or
Hartee-Fock spectrum at third order for nuclear saturation density
$n_0 = 0.17 \fmiq$ (corresponding to a Fermi momentum $k_F =
1.7\fmi$). We take this energy difference as a measure of convergence
for the potentials, as it includes both the uncertainty due to
different single-particle energies as well as the uncertainty in the
convergence of the many-body calculation.

\begin{table}[t]
\caption{Maximal energy difference between the second- and third-order 
contributions using a Hartree-Fock spectrum for the second-order and 
a free or Hartree-Fock spectrum for the third-order calculation at 
saturation density. Results are given for the different N$^3$LO NN 
potentials at the NN-only level and including the leading N$^2$LO 3N
forces with $\Lambda=2.0 \fmi$, $c_1=-0.75\gevi$ and $c_3=-4.77\gevi$.
The first three potentials exhibit a good convergence pattern with
both NN-only and including N$^2$LO 3N forces, and are therefore 
included in our complete N$^3$LO calculation. Note that the N$^3$LO
EGM 600/600 and 600/700 potentials will not be considered in our 
complete N$^3$LO calculation, because they have large $C_T$ couplings
(see the discussion of Table~\ref{tab:csct}).\label{tab:convergence}}
\begin{tabular*}{\columnwidth}{@{\extracolsep{\fill}}lcc}
\hline\hline
N$^3$LO NN potential & $|\Delta E^{(2/3)}_\text{NN-only}|$ 
& $|\Delta E^{(2/3)}_\text{NN/3N}|$ \\
\hline
EGM 450/500 MeV & $0.8\mev$ & $0.6\mev$ \\
EGM 450/700 MeV & $0.4\mev$ & $0.4\mev$ \\
EM 500 MeV      & $1.1\mev$ & $1.7\mev$ \\
\hline
EGM 550/600 MeV & $1.0\mev$ & $3.1\mev$ \\
EGM 600/600 MeV & $0.2\mev$ & $1.5\mev$ \\
EGM 600/700 MeV & $11.4\mev$& $16.1\mev$\\
EM 600 MeV      & $7.7\mev$ & $9.1\mev$ \\
\hline\hline
\end{tabular*}
\end{table}

At the NN level in Fig.~\ref{fig:NNconvergence}, the N$^3$LO EGM
potentials with cutoffs 450/500, 450/700, 550/600, and 600/600 MeV and
the N$^3$LO EM 500 MeV potential exhibit only small energy changes
from second to third order. The larger-cutoff potentials (N$^3$LO EGM
600/700 MeV and N$^3$LO EM 600 MeV), however, show large changes from
second to third order, as well as a large band for the range of
single-particle energies (especially for the EM 600 MeV
potential). This demonstrates that these potentials are
nonperturbative, see also Table \ref{tab:convergence}.

The convergence pattern is similar when the leading N$^2$LO 3N forces
are included. We show the results at this N$^3$LO NN and N$^2$LO 3N
level in Fig.~\ref{fig:NN3Nconvergence} for a 3N cutoff $\Lambda = 2.0
\fmi$ and a particular choice of $c_1 = -0.75\gevi$ and $c_3 =
-4.77\gevi$, although the general picture is unchanged for other
coupling values.

We find almost no change in the convergence pattern of the N$^3$LO EGM
450/500 and 450/700 MeV potentials; see Table~\ref{tab:convergence}.
This indicates that these potentials are perturbative for neutron
matter. For the N$^3$LO EGM 450/500 MeV potential, this is expected
already from the small Weinberg eigenvalues in Ref.~\cite{PPNP}, which
are a necessary condition for the perturbative convergence. The
perturbative convergence is a result of effective-range
effects~\cite{dEFT}, which weaken NN interactions at higher momenta,
combined with weaker tensor forces among neutrons, and with limited
phase space at finite density due to Pauli
blocking~\cite{nucmatt1}. For the EM 500 MeV potential the inclusion
of the N$^2$LO 3N forces decreases the uncertainty estimate from the
different single-particle energies, but increases the difference
between second and third order. This can be seen comparing
Figs.~\ref{fig:NNconvergence} and~\ref{fig:NN3Nconvergence} and is
reflected in the uncertainty estimate given in
Table~\ref{tab:convergence}. Since this potential is most commonly
used in nuclear structure calculations, we have decided to keep it in
our complete N$^3$LO calculation, in addition to the lower cutoff
N$^3$LO EGM 450/500 and 450/700 MeV potentials.

The N$^3$LO EGM 550/600 MeV potential is not used in the following
calculations because its uncertainty estimate (see
Table~\ref{tab:convergence}) increases by a factor of 3 when the
N$^2$LO 3N forces are included. This leads to a worse convergence
pattern compared to the low-cutoff EGM potentials. For the N$^3$LO EGM
600/700 MeV and EM 600 MeV potentials we find the situation unchanged
when including 3N forces and, thus, do not use these potentials for
the following calculations.  Even though the N$^3$LO EGM 600/600 MeV
potential exhibits a good convergence pattern, we will not use this
interaction because it breaks Wigner symmetry at the interaction level
(see the discussion of Table~\ref{tab:csct}).  Finally, we note that
our findings for the EM 500 MeV potential are consistent with
Ref.~\cite{Coraggio} (see Fig.~6 therein), where the authors studied
this potential at third order employing a Hartree-Fock spectrum.

\section{Results and discussion}
\label{sec:results}

Next, we present results using the EGM potentials with cutoffs 450/500
and 450/700 MeV and the EM 500 MeV potential. We discuss the
individual contributions first and then show the complete N$^3$LO results.

\begin{table*}[p]
\caption{Contributions from different N$^3$LO NN potentials and the
leading N$^2$LO 3N forces to the neutron-matter energy per particle
in MeV at nuclear saturation density. A Hartree-Fock spectrum
for the single-particle energies has been used. The 3N force is
for different $\Lambda$ in $\mathrm{fm}^{-1}$ and for different
$c_1$, $c_3$ in $\!\gevi$.\label{tab:higherorders}}
\begin{tabular*}{\textwidth}{@{\extracolsep{\fill}}lccccccccccc}
\hline\hline
NN potential & $c_1$/$c_3$ (3N) & $\Lambda$ & $E_\text{kin}^{(0)}$ & $E_\text{NN}^{(1)}$ & $E_\text{3N}^{(1)}$ & $E_1^{(2)}$ & $E_2^{(2)} + E_3^{(2)}$ & $E_4^{(2)}$ & $E_1^{(3)}$ & $E_2^{(3)} + E_3^{(3)}$ & $E_4^{(3)}$\\
\hline
EGM 450/500 MeV & $0$/$0$ & $-$ & $35.93$ & $-13.51$ & $0$ & $-7.88$ & $0$ & $0$ & $0.11$ & $0$ & $0$ \\
\cline{2-12}
& $-0.75$/$-4.77$ & $2.0$ & $35.93$ & $-13.51$ & $7.95$ & $-8.37$ & $-0.92$ & $-0.22$ & $0.14$ & $0.45$ & $0.03$\\
& & $2.5$ & $35.93$ & $-13.51$ & $9.06$ & $-7.94$ & $-3.41$ & $-0.95$ & $0.35$ & $0.32$ & $0.15$\\
\cline{2-12}
& $-1.13$/$-5.51$ & $2.0$ & $35.93$ & $-13.51$ & $9.37$ & $-8.47$ & $-1.08$ & $-0.31$ & $0.14$ & $0.55$ & $0.04$\\
& & $2.5$ & $35.93$ & $-13.51$ & $10.67$ & $-7.97$ & $-3.98$ & $-1.31$ & $0.39$ & $0.81$ & $0.22$\\
\hline
EGM 450/700 MeV & $0$/$0$ & $-$ & $35.93$ & $-19.39$ & $0$ & $-4.49$ & $0$ & $0$ & $0.08$ & $0$ & $0$ \\
\cline{2-12}
& $-0.75$/$-4.77$ & $2.0$ & $35.93$ & $-19.39$ & $7.95$ & $-4.77$ & $-0.63$ & $-0.22$ & $0.09$ & $0.25$ & $0.02$\\
&  & $2.5$ & $35.93$ & $-19.39$ & $9.06$ & $-4.52$ & $-2.45$ & $-0.96$ & $0.19$ & $0.40$ & $0.13$\\
\cline{2-12}
& $-1.13$/$-5.51$ & $2.0$ & $35.93$ & $-19.39$ & $9.37$ & $-4.83$ & $-0.74$ & $-0.31$ & $0.10$ & $0.31$ & $0.03$\\
&  & $2.5$ & $35.93$ & $-19.39$ & $10.67$ & $-4.54$ & $-2.87$ & $-1.32$ & $0.21$ & $0.50$ & $0.19$\\
\hline
EM 500 MeV & $0$/$0$ & $-$ & $35.93$ & $-17.49$ & $0$ & $-6.71$ & $0$ & $0$ & $1.13$ & $0$ & $0$ \\
\cline{2-12}
& $-0.75$/$-4.77$ & $2.0$ & $35.93$ & $-17.49$ & $7.95$ & $-7.13$ & $-0.52$ & $-0.19$ & $1.26$ & $0.39$ & $0.02$\\
&  & $2.5$ & $35.93$ & $-17.49$ & $9.06$ & $-6.84$ & $-2.27$ & $-0.83$ & $1.21$ & $0.96$ & $0.14$\\
\cline{2-12}
& $-1.13$/$-5.51$ & $2.0$ & $35.93$ & $-17.49$ & $9.37$ & $-7.21$ & $-0.61$ & $-0.27$ & $1.29$ & $0.47$ & $0.03$\\
& & $2.5$ & $35.93$ & $-17.49$ & $10.67$ & $-6.87$ & $-2.67$ & $-1.14$ & $1.23$ & $1.17$ & $0.20$\\
\hline\hline
\end{tabular*}
 
\end{table*}
\begin{figure*}[p]
\begin{center}
\includegraphics[height=5.2cm,clip=]{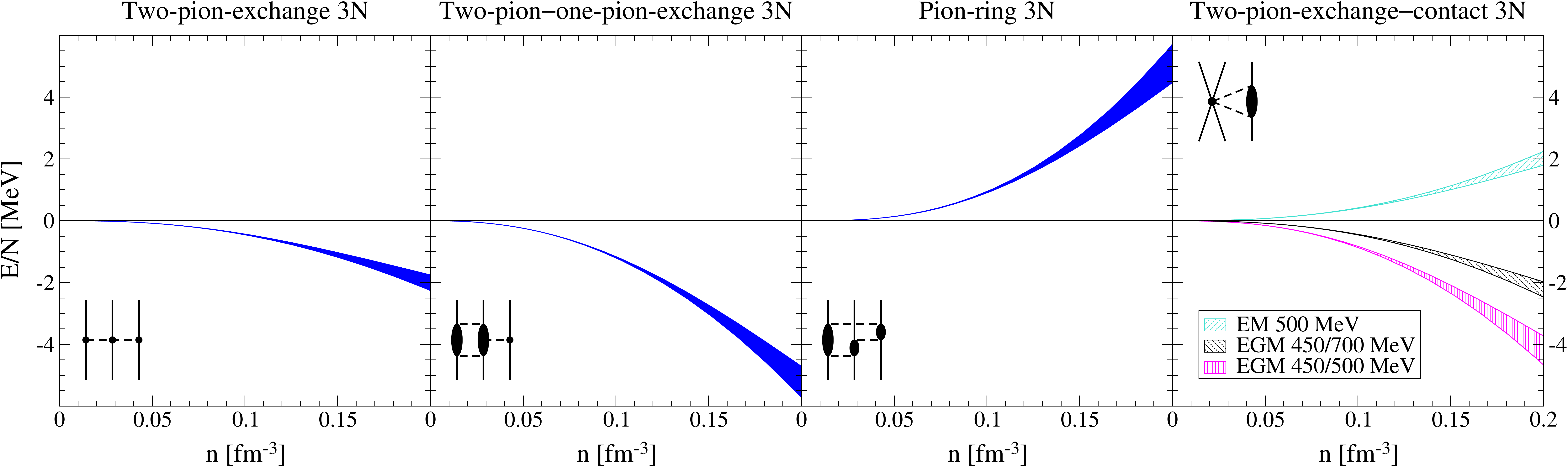}
\newline\newline
\includegraphics[height=5.3cm,clip=]{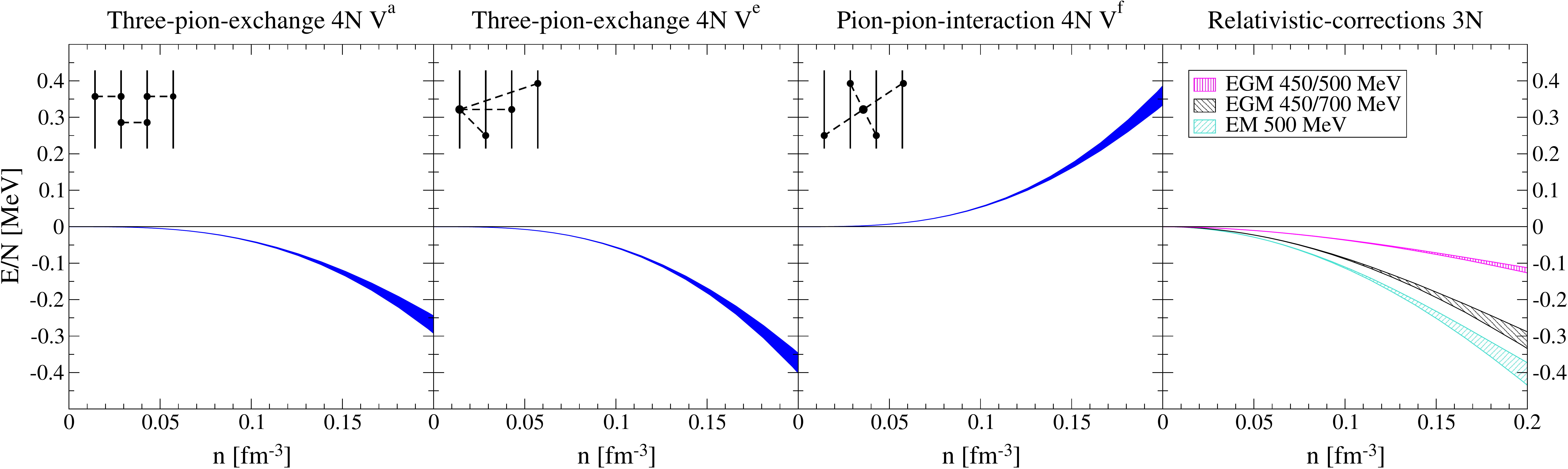}
\end{center}
\caption{(Color online) Energy per particle as a function of density
for all individual N$^3$LO 3N- and 4N-force contributions to neutron
matter at the Hartree-Fock level. All bands
are obtained by varying the 3N/4N cutoffs $\Lambda = 2-2.5 \fmi$. For the 
two-pion-exchange--contact and the relativistic-corrections 3N forces,
the different bands correspond to the different NN contacts, $C_T$ 
and $C_S$, determined consistently for the N$^3$LO EM/EGM potentials.
The inset diagram illustrates the 3N/4N force topology of the particular
contributions.
\label{fig:individualnm}}
\end{figure*}

\begin{figure*}[t]
\begin{center}
\includegraphics[width=0.975\textwidth,clip=]{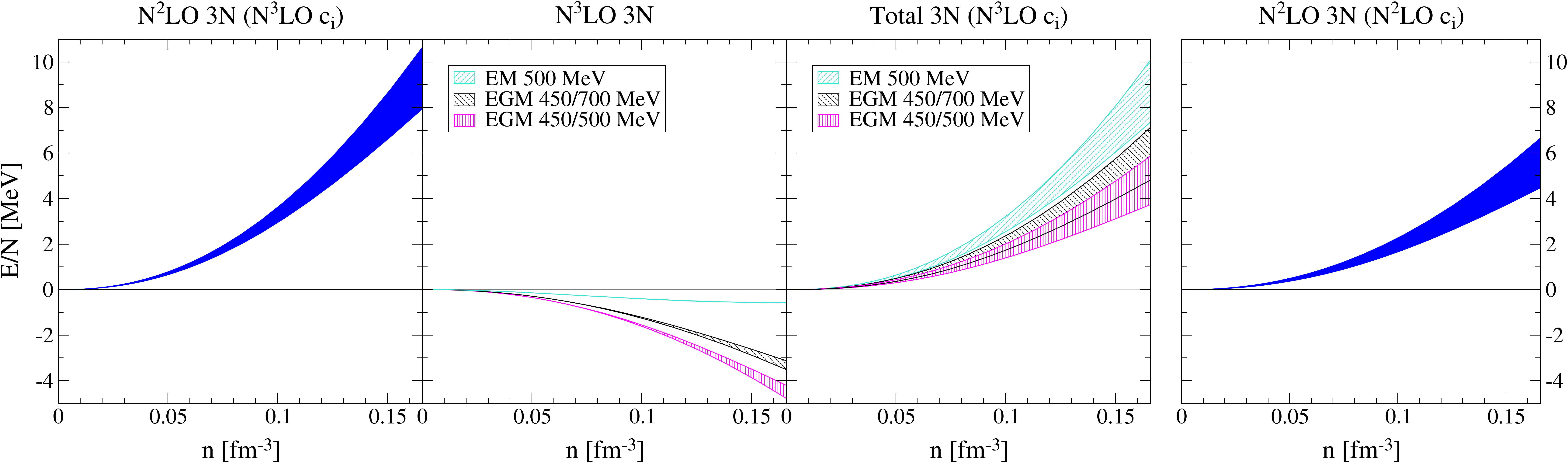}
\end{center}
\caption{(Color online) Contributions from 3N forces in the Hartree-Fock
approximation at N$^2$LO plus N$^3$LO (first three panels) in comparison
with the 3N contribution in a N$^2$LO calculation (fourth panel). The 
first panel shows the N$^2$LO 3N contribution in the N$^3$LO calculation,
using the N$^3$LO values of the $c_i$ couplings, and the second panel
gives the N$^3$LO 3N contribution. The third panel shows the total 3N
contribution at N$^3$LO (the sum of the first two panels). This is
compared in the fourth panel to the 3N contribution at N$^2$LO, using
the $c_i$ values recommended for an N$^2$LO calculation (see 
Table~\ref{tab:c_i}). For the EGM potentials the total 3N contribution
at N$^3$LO differs by less than $1 \mev$ compared to the N$^2$LO
results. However, for the EM potential, the result changes by almost
$3 \mev$. All bands include the $c_i$ range from Table~\ref{tab:c_i}
and the 3N cutoff variation.\label{fig:3Ncomparison}}
\end{figure*}

\subsection{N$^\text{3}$LO NN and N$^\text{2}$LO 3N forces}

The N$^3$LO NN and N$^2$LO 3N forces have been evaluated at the
Hartree-Fock level and including second- and third-order
contributions. Beyond Hartree Fock, N$^2$LO 3N forces are taken into
account as density-dependent two-body interactions~\cite{nm}. The
kinetic energy, Hartree-Fock, and individual higher-order interaction
contributions for the N$^3$LO NN and N$^2$LO 3N parts are given in
Table~\ref{tab:higherorders} for different values of $\Lambda$ and the
$c_i$ couplings. Vanishing $c_i$ in the 3N forces correspond to NN
forces only. Table~\ref{tab:higherorders} shows that the dominant
higher-order contributions are due to the second-order NN-NN part
$E^{(2)}_1$. The second-order NN-3N parts $E^{(2)}_2 + E^{(2)}_3$ are
of the order of $1 \mev$ and only larger for the large 3N cutoff. All
higher-order contributions with 3N forces are systematically
smaller. We emphasize that the N$^2$LO 3N contributions beyond Hartree
Fock are larger than in Ref.~\cite{nm}, and therefore also the
many-body calculation converges more slowly, because the N$^2$LO 3N forces are
stronger due to the large N$^3$LO values of the $c_i$ couplings.

The NN-only energies per particle are $14.7$, $12.1$, and $12.9 \mev$
at saturation density for the EGM 450/500, EGM 450/700 MeV, and EM
500 MeV N$^3$LO potentials, respectively. Inclusion of 3N forces at
N$^2$LO adds another $7 \pm 1.5\mev$ per particle at saturation
density (using the larger N$^3$LO $c_i$ values, see Table~\ref{tab:c_i}).

\subsection{N$^\text{3}$LO 3N and 4N forces}

The N$^3$LO many-body forces have been evaluated in the Hartree-Fock
approximation. We have not calculated higher-order contributions
because of their involved structure. The Hartree-Fock approximation is
expected to be reliable based on the findings of Ref.~\cite{nm}. In
addition, higher-order contributions with N$^3$LO many-body forces are
not enhanced by large $c_i$ couplings, and the N$^3$LO many-body
forces are smaller than at N$^2$LO, leading to smaller higher-order
corrections.

We show the individual contributions of the 3N and 4N forces in
Fig.~\ref{fig:individualnm}. The bands correspond to the cutoff
variation $\Lambda=2 - 2.5 \fmi$. In the shorter-range
two-pion-exchange--contact and the relativistic corrections 3N forces,
three different bands are shown. These correspond to the different NN
contacts, $C_T$ and $C_S$, determined consistently for the
different N$^3$LO EM/EGM potentials.

The two-pion-exchange 3N forces at N$^3$LO yield an energy per
particle of $-1.5 \mev$ at saturation density, which is $\sim 1/3$ of
the 3N contributions at N$^2$LO and sets the natural scale. The
two-pion--one-pion-exchange and the pion-ring 3N forces lead to
relatively large contributions of $-3.5 \mev$ and $+3.3 \mev$ per
particle at $n_0$, respectively. The contributions of the
two-pion-exchange--contact 3N forces range between $-2.8 \mev$ and
$+1.3 \mev$ per particle at $n_0$, depending on the NN potential. In
the topologies with relatively large expectation values, the large
$c_i$ couplings will enter in many-body forces at N$^4$LO~\cite{Krebs}.
This may reflect important $\Delta$ contributions shifted to N$^4$LO, as
discussed above. Finally, the relativistic corrections contribute
$-(0.1-0.3) \mev$ to the energy per particle at $n_0$ and are small
compared with the other topologies.

As shown in Fig.~\ref{fig:3Ncomparison} (second panel) the sum of the
N$^3$LO 3N contributions yields an energy of $-(3-5) \mev$ per
particle at saturation density for the EGM potentials and a small
contribution of $-0.5 \mev$ for the EM potential. This shows that the
N$^3$LO 3N contribution can be significant, compared to the N$^2$LO 3N
energy of $7 \pm 1.5\mev$ per particle (note that the first panel of
Fig.~\ref{fig:3Ncomparison} only gives this contribution at the
Hartree-Fock level). The relatively large N$^3$LO 3N contributions are
compensated by the larger N$^3$LO $c_i$ values, entering the 3N force
at N$^2$LO. This can be seen in Fig.~\ref{fig:3Ncomparison} where the
total 3N contribution at N$^3$LO (third panel) is compared at the
Hartree-Fock level to the 3N contribution at N$^2$LO (fourth panel),
which uses the $c_i$ values recommended for an N$^2$LO calculation
(see Table~\ref{tab:c_i}). For the EGM potentials the total 3N
contribution changes by less than $1 \mev$ going from N$^2$LO to
N$^3$LO. Because the N$^3$LO 3N contribution is small for the EM
potential, this results in a difference of about $3 \mev$ when going
from N$^2$LO to N$^3$LO for the EM case, due to the modified $c_i$
couplings at N$^3$LO.

Only three N$^3$LO 4N topologies give nonvanishing contributions to
neutron matter. We show their results in Fig.~\ref{fig:individualnm}.
The two three-pion-exchange diagrams $V^a$ and $V^e$ are attractive
with energies of $-0.16\mev$ and $-0.25\mev$ per particle at
saturation density. The pion-pion-interaction 4N forces ($V^f$) are
repulsive with $0.22 \mev$ per particle at $n_0$. The latter two
diagrams almost cancel each other, such that the total contribution of
the leading 4N forces is about $-0.18 \mev$ per particle at
$n_0$. However, also for the 4N forces additional larger contributions
from $\Delta$ excitations may arise at N$^4$LO~\cite{Kaiser}.

At the Hartree-Fock level, the 3N/4N contributions change by less than
5\% if the cutoff is taken to infinity (i.e., $f_R = 1$). However,
since we also include N$^2$LO 3N forces beyond Hartree Fock, a
consistent regulator is required. Finally, we compare our 4N results
with those of Refs.~\cite{Fiorilla2011,Kaiser}, which considered only
the 4N interactions $V^e$ and $V^f$ and found their sum to be about
$-11\kev$ per particle at $n_0$. This is in agreement with
our results, if we take $f_R = 1$ as in Refs.~\cite{Fiorilla2011,Kaiser}.

\subsection{Complete calculation at N$^\text{3}$LO}

\begin{figure}[t]
\begin{center}
\includegraphics[width=0.9\columnwidth,clip=]{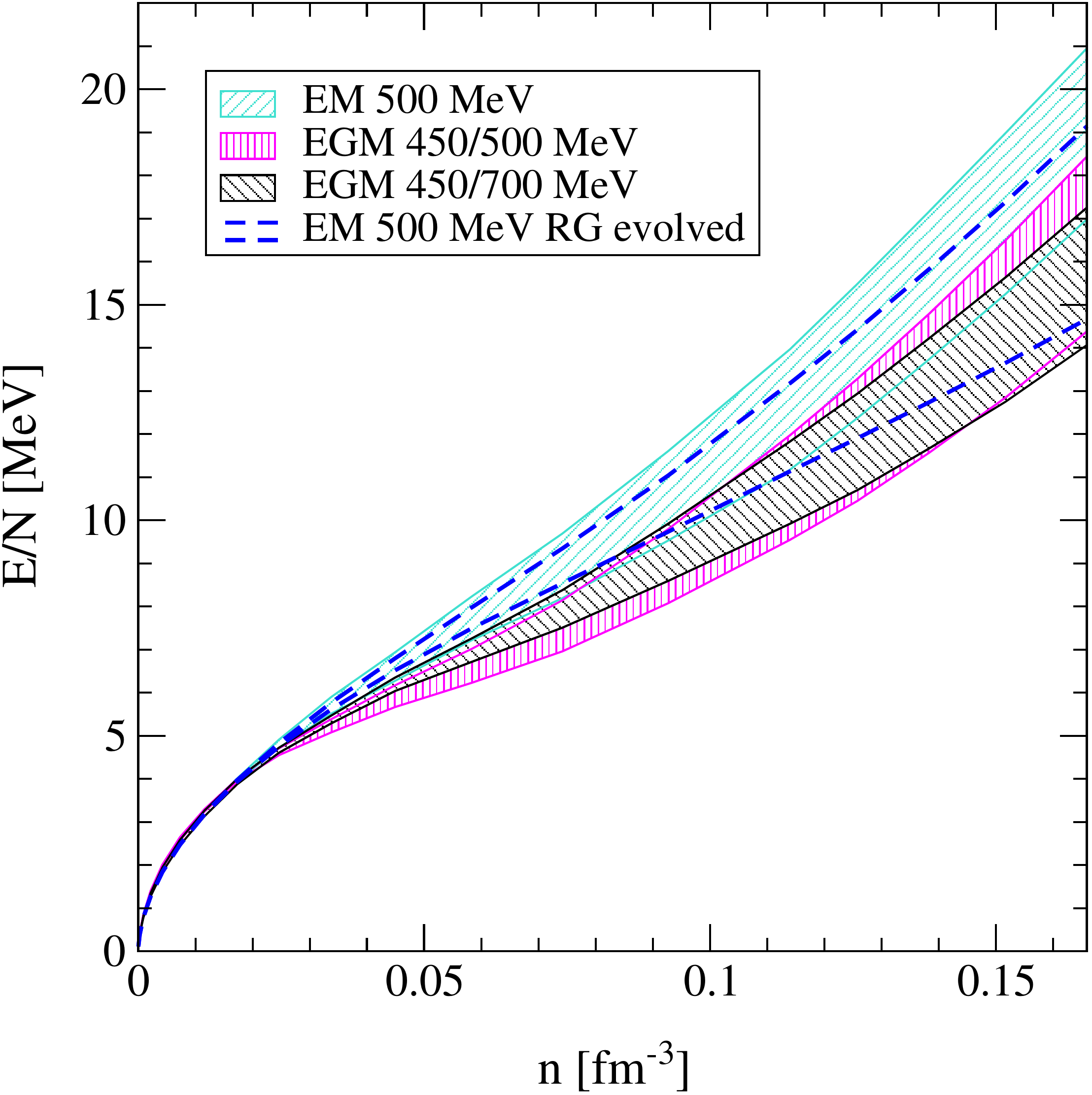}
\end{center}
\caption{(Color online) Neutron-matter energy per particle as a
function of density including NN, 3N, and 4N forces to N$^3$LO. The
three overlapping bands are labeled by the different NN potentials and
include uncertainty estimates due to the many-body calculation, the
low-energy $c_i$ constants, and by varying the 3N/4N cutoffs (see text
for details). For comparison, we show the results for the RG-evolved
NN EM 500 MeV potential including only N$^2$LO 3N forces from
Ref.~\cite{nm}.\label{fig:fullN3LO}}
\end{figure}

The complete N$^3$LO result for neutron matter is shown in
Fig.~\ref{fig:fullN3LO}, which includes all many-body interactions to
N$^3$LO~\cite{fullN3LO}. For all shown potentials the uncertainties in
the $c_i$ couplings dominate the width of the bands (compare to the
bands in the upper row of Fig.~\ref{fig:NN3Nconvergence}).

\color{black}
At saturation density, we obtain for the
energy per particle
\begin{equation}
\frac{E}{N}(n_0) = 14.1-21.0 \mev \,.
\end{equation}
This range is based on different NN potentials, a variation of the
couplings $c_1 = -(0.75 - 1.13)\gevi$ and $c_3 = -(4.77 - 5.51) \gevi$,
and on the 3N/4N-cutoff variation $\Lambda = 2 - 2.5\fmi$. In
addition, the uncertainty in the many-body calculation is included,
as discussed above.

As shown in Fig.~\ref{fig:fullN3LO}, our results are consistent with
previous calculations based on RG-evolved NN interactions at N$^3$LO
and 3N interactions at N$^2$LO~\cite{nm}. These calculations adopted a
conservative $c_i$ range but are based on the EM 500 MeV NN potential
only, which results in a narrower band compared to the N$^3$LO band.
In Ref.~\cite{fullN3LO}, we compared our results to calculations based
on lattice EFT~\cite{NLOlattice} and quantum Monte Carlo at low
densities~\cite{GC}, as well as to variational methods~\cite{APR} and
auxiliary field diffusion Monte Carlo~\cite{GCR} based on
phenomenological NN and 3N potentials, and found that they are also
consistent with the N$^3$LO band. However, the latter calculations do
not provide theoretical uncertainties.

\begin{figure}[t]
\begin{center}
\includegraphics[width=0.9\columnwidth,clip=]{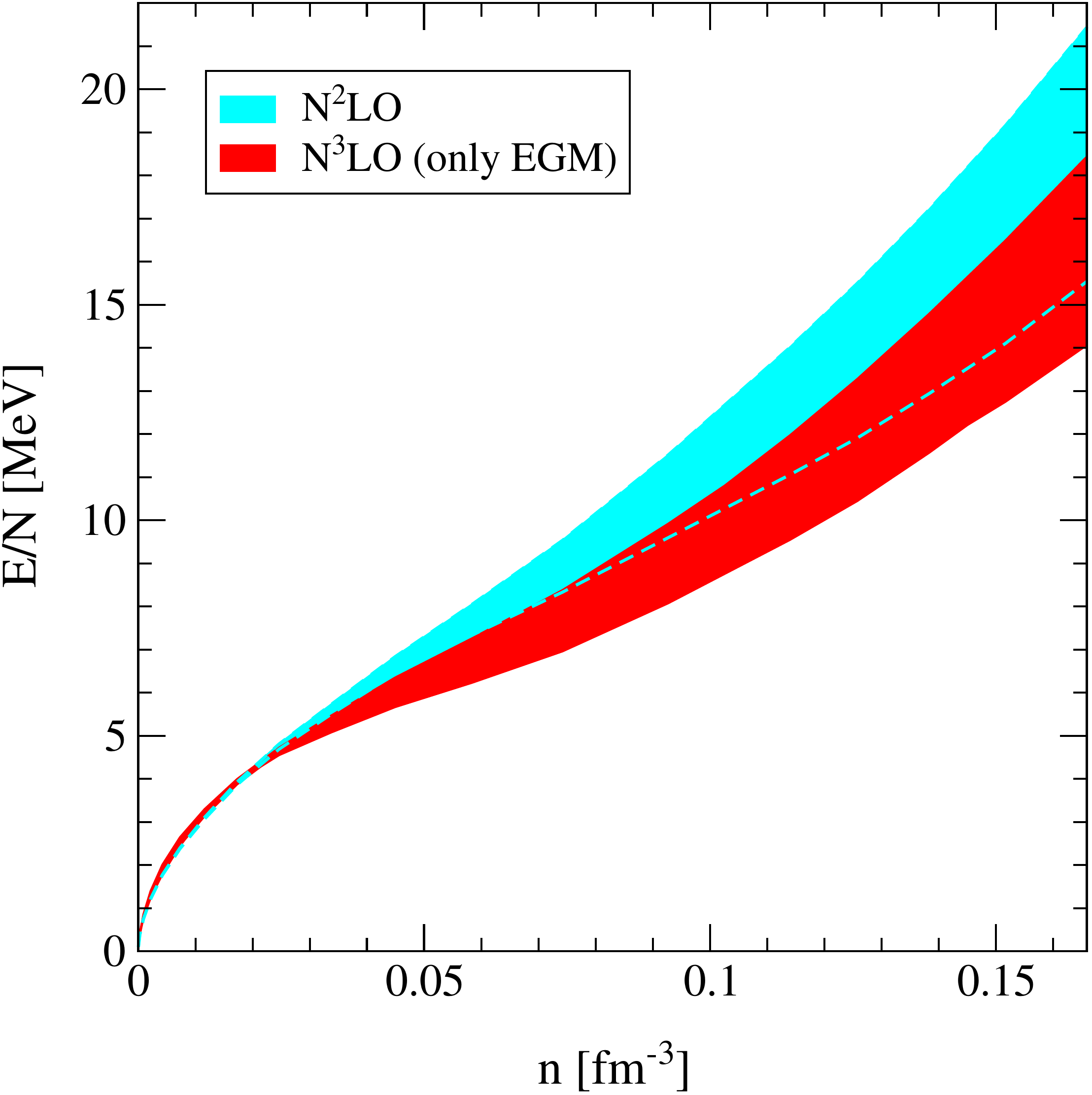}
\end{center}
\caption{(Color online) Neutron-matter energy per particle as a 
function of density at N$^2$LO (upper blue band that extends to the
dashed line) and N$^3$LO (lower red band). The bands are based on the
EGM NN potentials and include uncertainty estimates as in
Fig.~\ref{fig:fullN3LO}.\label{fig:comparison}}
\end{figure}

In Fig.~\ref{fig:comparison} we compare the convergence from N$^2$LO
to N$^3$LO in the same calculational setup. For this comparison, we
consider only the EGM potentials with cutoffs 450/500 and 450/700 MeV,
since no EM N$^2$LO potential is available. This leads to an N$^3$LO
energy range of $14.1 - 18.4\mev$ per particle at $n_0$. For the
N$^2$LO band in Fig.~\ref{fig:comparison}, we have estimated the
theoretical uncertainties in the same way and found an energy of $15.5
- 21.4\mev$ per particle at $n_0$. The two bands overlap but the range
of the band is reduced only by a factor of $2/3$, which is larger than
the $1/3$ expected from the EFT power counting. We attribute this to
$\Delta$ effects (as discussed above). This can be improved by
including the $\Delta$ in chiral EFT explicitly or by going to
N$^4$LO~\cite{Krebs}.

Finally, it is important to construct NN potentials at N$^2$LO and
N$^3$LO covering the range of the $c_i$ values. At N$^3$LO, we expect
that the differences in the $c_i$ can be absorbed partly by $Q^4$
contact interactions in the fits to NN scattering. In addition, the
many-body-calculation uncertainties can be reduced further by
including the N$^3$LO many-body forces beyond the Hartree-Fock level.

\section{Applications}
\label{sec:applications}

\subsection{Symmetry energy and its density derivative}

The symmetry energy $S_v$ and its density derivative $L$ provide
important input for astrophysics~\cite{LL}. To calculate these, we need to
extend the neutron-matter energy to asymmetric matter. For the energy
per particle $\epsilon$, we follow Ref.~\cite{nstar_long} and take an
expression that includes kinetic energy plus interaction energy that
is quadratic in the neutron excess $1-2x$, where $x$ is the proton
fraction,
\begin{align}
\epsilon(\bar{n},x) &= T_0 \biggl[ \frac{3}{5} \Bigl[x^{\frac{5}{3}}
+(1-x)^{\frac{5}{3}} \Bigr] (2\bar{n})^{\frac{2}{3}} \nonumber\\[1mm]
&\quad\quad\:\: -\bigl[(2\alpha-4\alpha_L)x(1-x)+\alpha_L \bigr]
\bar{n} \nonumber\\
&\quad\quad\:\: +\bigl[(2\eta-4\eta_L)x(1-x)+\eta_L \bigr]
\bar{n}^{\frac{4}{3}} \biggr] \,,
\label{extension}
\end{align}
where $\bar{n}=n/n_0$ and $T_0=(3\pi^2n_0/2)^{2/3}/(2m) = 36.84\mev$
is the Fermi energy of symmetric nuclear matter at saturation density.
The parameters $\alpha = 5.87$ and $\eta=3.81$ are determined through
fits to the empirical saturation point of nuclear matter, and
$\alpha_L$ and $\eta_L$ through fits to the neutron-matter results of
Fig.~\ref{fig:fullN3LO} (for details on this strategy, see
Ref.~\cite{nstar_long}). Equation~(\ref{extension}) provides very good
fits to the N$^3$LO energy band.

\begin{table}[b]
\vspace*{-3mm}
\caption{Ranges for the symmetry energy $S_v$ and its density derivative
$L$ at nuclear saturation density.\label{tab:svl}}
\begin{tabular*}{\columnwidth}{@{\extracolsep{\fill}}lc}
\hline\hline
& range \quad \\
\hline
Symmetry energy $S_v(n_0)$ & $28.9 - 34.9\mev$ \quad \\
Density derivative $L(n_0)$ & $43.0 - 66.6\mev$ \quad \\
\hline\hline
\end{tabular*}
\end{table}]

We can then calculate the symmetry energy
\begin{equation}
S_v(n) = \frac{1}{8} \frac{\partial^2 \epsilon(\bar{n},x)}{\partial x^2} 
\biggr|_{\bar{n}=1, x=1/2} \,,
\end{equation}
and its density derivative
\begin{equation}
L(n) = \frac{3}{8} \frac{\partial^3 \epsilon(\bar{n},x)}{\partial \bar{n} 
\partial x^2} \biggr|_{\bar{n}=1, x=1/2} \,.
\end{equation}
The $L$ parameter basically determines the pressure of neutron matter.
In addition, because the expression~(\ref{extension}) is fit to the
empirical saturation point (with small uncertainties), the symmetry
energy and its density derivative at $n_0$ and their theoretical
uncertainties are essentially determined by the neutron-matter results.

The predicted ranges for $S_v$ and $L$ at saturation density are given
in Table~\ref{tab:svl}. In Ref.~\cite{fullN3LO}, we have shown that
$S_v$ and $L$ are also correlated and overlap with the results for
RG-evolved NN interactions with N$^2$LO 3N forces~\cite{LL,nstar_long},
but, due to the additional density dependencies from N$^3$LO many-body
forces, this correlation is not as tight. The $S_v$ and $L$ ranges are
also in very good agreement with experimental constraints from nuclear
masses~\cite{Kortelainen2010} and from the dipole polarizability of
$^{208}$Pb~\cite{Tamii2011} (see also Refs.~\cite{LL,fullN3LO}).

\subsection{Constraints for supernova equations of state and neutron stars}

\begin{figure}[t]
\begin{center}
\includegraphics[height=0.9\columnwidth,clip=]{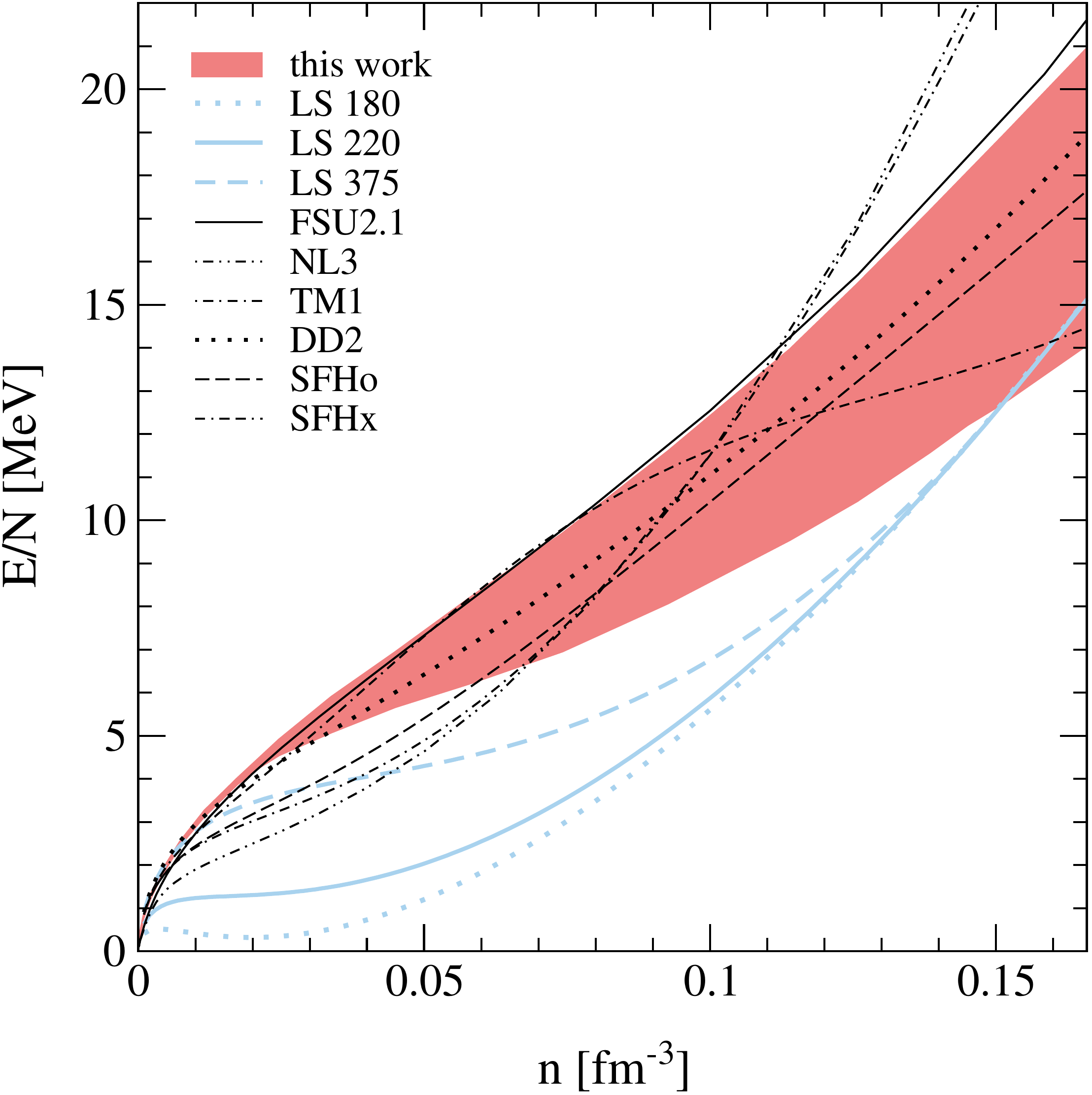}
\end{center}
\caption{(Color online) Comparison of the neutron-matter energy at 
N$^3$LO of Fig.~\ref{fig:fullN3LO} (red band) with equations of 
state for core-collapse supernova simulations provided by Lattimer-Swesty
(LS~\cite{Lattimer1991} with different incompressibilities, 180, 220,
and $375 \mev$), G.~Shen (FSU2.1, NL3 \cite{GShenprivate}), Hempel
(TM1, SFHo, SFHx \cite{Hempelprivate}), and Typel (DD2
\cite{Typelprivate}).\label{fig:EOScomparison}}
\end{figure}

The neutron-matter results also provide constraints for the nuclear
equation of state. Here we focus on comparisons to equations of state
for core-collapse supernova simulations. In Fig.~\ref{fig:EOScomparison},
we compare the N$^3$LO neutron-matter band (red band) to the
Lattimer-Swesty (LS) equation of state~\cite{Lattimer1991}, which is
most commonly used in simulations, and to different relativistic
mean-field-theory equations of state based on the density functionals
DD2~\cite{Typelprivate}, FSU2.1~\cite{GShen2011oc}, NL3~\cite{GShen2011t},
SFHo, SFHx~\cite{Steiner2012}, and TM1~\cite{HShen2011}. At low
densities only the DD2, FSU2.1, and SFHx equations of state are
consistent with the N$^3$LO neutron-matter band. The other supernova
equations of state underestimate the energy for densities below $\sim
0.5 n_0$ and even at higher density in the LS cases. This density
range covers the outer regions of the (proto-) neutron star, where also
protons, nuclei, and electrons are relevant. Nevertheless, the
deficiencies in the nuclear interactions of these equations of state
will also affect the chemical potentials and the neutrino
response. Around saturation density, the LS and SFHo equations of
state become consistent with the N$^3$LO band. We also find that 
the NL3 and TM1 equations of state have a too strong density dependence,
which leads to unnaturally large $S_v$ and $L$ values. In addition,
Fig.~\ref{fig:EOScomparison} exhibits a strange density dependence of SFHx.

\begin{figure}[t]
\begin{center}
\includegraphics[height=0.9\columnwidth,clip=]{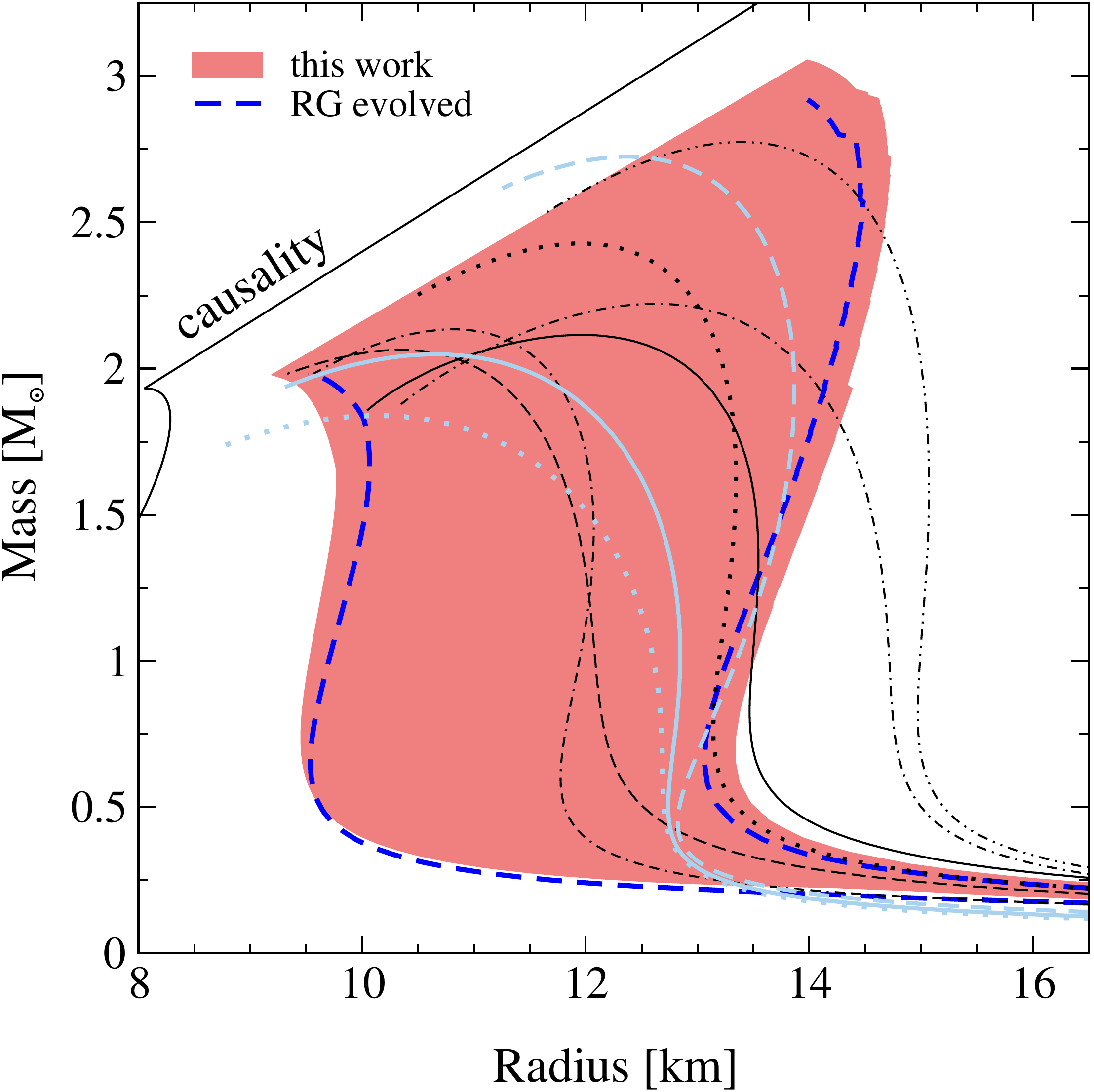}
\end{center}
\caption{(Color online) Constraints on the mass-radius diagram of 
neutron stars based on our neutron-matter results at N$^3$LO following
Ref.~\cite{nstar_long} for the extension to neutron-star matter and to
high densities (red band), in comparison to the constraints from
calculations based on RG-evolved NN interactions (thick dashed blue
lines)~\cite{nstar_long}. We also show the mass-radius relations
obtained from the equations of state for core-collapse supernova
simulations shown in Fig.~\ref{fig:EOScomparison}~\cite{ConnorKleiner,%
Lattimer1991,GShen2011t,GShen2011oc,HShen2011,Steiner2012,Hempel2012}.
The legend for the thin lines is as described in the caption to Fig.~\ref{fig:EOScomparison}.
\label{fig:nstar}}
\end{figure}

Next, we use the N$^3$LO neutron-matter results to provide constraints
for the structure of neutron stars. We follow Ref.~\cite{nstar_long}
for incorporating $\beta$ equilibrium and for the extension to high
densities using piecewise polytropes that are constrained by causality
and by the requirement to support a $1.97 \pm 0.04 \, M_\odot$ neutron
star~\cite{Demorest}, the heaviest precisely measured neutron star to
date. The resulting constraints on the neutron star mass-radius
diagram are shown in Fig.~\ref{fig:nstar} by the red band. This band
represents an envelope of a large number of individual equations of
state reflecting the uncertainties in the N$^3$LO neutron-matter calculation
and in the polytropic extensions to high densities~\cite{nstar_long}.
Figure~\ref{fig:nstar} confirms the predicted radius range of
Ref.~\cite{nstar_long} of $9.7-13.9 \km$ for a $1.4 \, M_\odot$
neutron star. The largest supported neutron star mass is found to be
$3.1 \, M_\odot$, with a corresponding radius of about $14 \km$. We
also find very good agreement with the mass-radius constraints from
the neutron-matter calculations based on RG-evolved NN interactions
with N$^2$LO 3N forces~\cite{nstar_long}, which are shown by the thick
dashed blue lines in Fig.~\ref{fig:nstar}.

\begin{figure*}[t]
\begin{center}
\includegraphics[width=\textwidth]{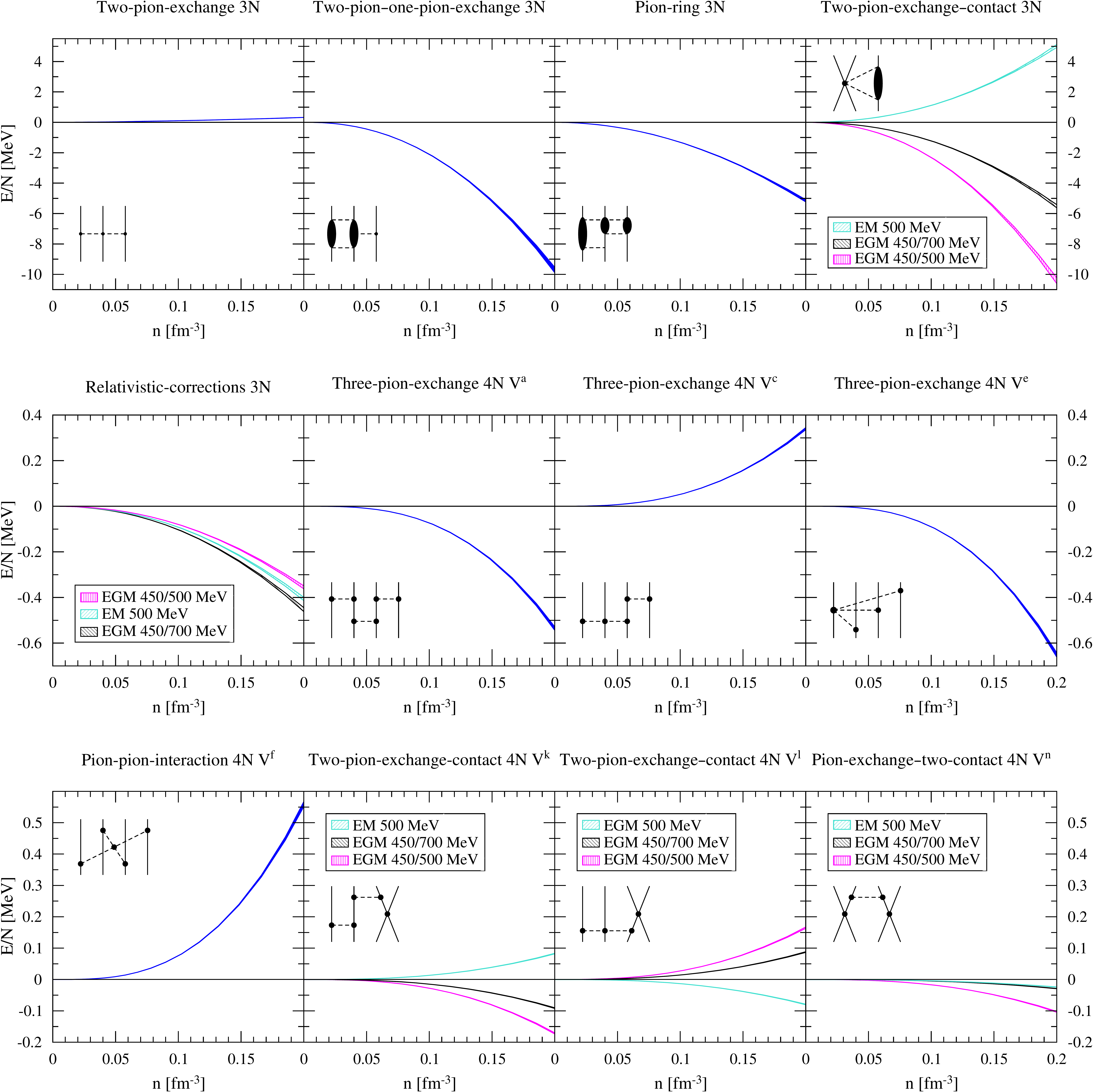}
\end{center}
\caption{(Color online) Energy per particle versus density for all 
individual N$^3$LO 3N- and 4N-force contributions to symmetric nuclear
matter at the Hartree-Fock level. All bands are obtained by varying
the 3N/4N cutoff $\Lambda = 2-2.5 \fmi$. For the
two-pion-exchange--contact, the relativistic-corrections 3N forces,
and the short-range 4N forces, the different bands correspond to the
different NN contacts, $C_T$ and $C_S$, determined consistently for
the N$^3$LO EM/EGM potentials. The inset diagram illustrates the 3N/4N
force topology of the particular contribution.\label{fig:individualsym}}
\end{figure*}

In addition, we show in Fig.~\ref{fig:nstar} the mass-radius relations
obtained from equations of state for core-collapse supernova
simulations~\cite{ConnorKleiner,Lattimer1991,GShen2011t,GShen2011oc,%
HShen2011,Steiner2012,Hempel2012}. The inconsistency in
Fig.~\ref{fig:EOScomparison} of many of the equations of state with
the N$^3$LO neutron-matter band at low densities results in a large
spread of very low mass/large radius neutron stars, where the red band
is considerably narrower in Fig.~\ref{fig:nstar} (note that the red
band includes a standard crust equation of state below $0.5 \,
n_0$~\cite{nstar_long}). For typical neutron stars, our calculations
rule out the NL3 and TM1 equations of state, which produce too-large
radii. Finally, we emphasize that these constraints not only are
important for neutron star structure and for the supernova equation of
state but also provide nuclear physics constraints for the
gravitational wave signal in neutron star
mergers~\cite{Bauswein1,Bauswein2}.

\section{First estimate for symmetric nuclear matter}
\label{sec:symmetric}

We present first results for the N$^3$LO many-body forces in symmetric
nuclear matter in the Hartree-Fock approximation. However, we
emphasize that these results should be considered as a preview and to
show their importance, because it is crucial to include contributions
beyond the Hartree-Fock level~\cite{nucmatt2}. Such calculations can
also be facilitated by a similarity RG evolution of NN and 3N
forces~\cite{3Nevolution,nm_3Nevolution} in order to improve the
convergence of the many-body calculation.

The energy per particle of symmetric matter is evaluated as in
Sec.~\ref{sec:mbdetailsHF} summing also over both isospin states
[see Eq.~(\ref{eq:snm_me})]. In Appendix~\ref{app:sym}, the
expressions for the N$^3$LO 3N- and 4N-interaction matrix elements are
given in detail. Our results for the individual contributions from
N$^3$LO many-body forces are shown in Fig.~\ref{fig:individualsym}.
Compared to the neutron-matter results, the individual contributions
are larger in magnitude in symmetric matter, requiring calculations
beyond the Hartree-Fock level. However, the bands from cutoff
variation are narrower, because the Fermi momentum corresponding to
saturation density is lower in symmetric matter.

For the two-pion-exchange N$^3$LO 3N forces the energy is small, with
$0.24 \mev$ per particle at $n_0$ due to cancellations among the
individual parts in symmetric matter. The other 3N topologies are
large and attractive: the two-pion--one-pion-exchange and the
pion-ring 3N interactions give energies of $-6.5 \mev$ and $-3.6 \mev$
per particle at $n_0$, respectively. The contribution of the
two-pion-exchange--contact 3N interaction ranges from $-7.0 \mev$ to
$+3.4 \mev$, depending on the NN potential. As expected from our
neutron-matter results, the large 3N contributions in these topologies
can be attributed to the physics from $\Delta$ excitations, which will
lead to large $c_i$ contributions at N$^4$LO in these topologies (or
at N$^3$LO in $\Delta$-full chiral EFT). As in neutron matter, the
contributions from relativistic-corrections 3N forces are small with
$-(0.24-0.39) \mev$ per particle at $n_0$.

Since nuclear saturation is a result of cancellation effects of large
energy contributions~\cite{nucmatt2}, the increased strengths of the
$c_i$ couplings at N$^3$LO compared to N$^2$LO is expected to play an
important role for predictions of symmetric matter. Furthermore, in
contrast to neutron matter, we find that the total N$^3$LO 3N
contribution at the Hartree-Fock level depends more strongly on the NN
potentials used: For the EM $500 \mev$ potential, we find $-7 \mev$
per particle at $n_0$, whereas for the EGM potentials, we find
$-(15-17) \mev$. To understand this better, improved NN potential fits
(following Ref.~\cite{Ekstroem}) and also those for different $c_i$
couplings will be important. These N$^3$LO energies should be compared
with a total N$^2$LO 3N energy at the Hartree-Fock level of the order
of $15 \mev$ per particle at $n_0$, using the large N$^3$LO $c_i$
values (see Table~\ref{tab:c_i} and accordingly chosen $c_4 = 3.34 -
3.71 \gevi$~\cite{Krebs}) and typical $c_D$, $c_E$ values~\cite{nucmatt1}.
All these findings show that including N$^3$LO 3N contributions
beyond the Hartree-Fock level will be crucial.

Figure~\ref{fig:individualsym} also shows our results for the
individual N$^3$LO 4N-force contributions in symmetric matter. The
long-range three-pion-exchange 4N interactions $V^a$ and $V^e$ are
attractive with energies $-0.32 \mev$ and $-0.39 \mev$ per particle at
$n_0$, respectively; while the $V^c$ interaction is repulsive with
$0.21\mev$ per particle at $n_0$. The pion-pion-interaction 4N force
$V^f$ also gives a repulsive contribution of $0.33\mev$ per particle
at $n_0$. The shorter-range parts $V^k$, $V^l$, and $V^n$ contain one
or two spin-dependent NN contact interactions and depend on $C_T$. The
two midrange topologies involving only one NN contact ($V^k$ and
$V^l$) almost cancel against each other ($-0.11$ to $+0.05 \mev$ per
particle at $n_0$, depending on the NN potential, and $-0.05$ to
$+0.10 \mev$, respectively). The shortest-range topology with two NN
contacts ($V^n$) contributes even less ($-0.06$ to $-0.01 \mev$ per
particle at $n_0$). In total, the leading 4N forces give an attractive
contribution of $-(0.18-0.23) \mev$ per particle at $n_0$, with a
strong density dependence $\sim n^3$.

As a check, we can compare our results to the studies of the $V^e$ and
$V^f$ 4N forces of Refs.~\cite{Fiorilla2011,Kaiser}, which obtained a
contribution to the energy per particle of $-53\kev$ at $n_0$. This is
in agreement with our result for the sum of these two topologies:
$-(56 \pm 2) \kev$, where the small difference is due to $f_R=1$ in
Refs.~\cite{Fiorilla2011,Kaiser}. So far only the leading 4N forces
have been derived completely. Recently, Kaiser studied $\Delta$
contributions to 4N forces~\cite{Kaiser}, which enter at N$^4$LO in
$\Delta$-less chiral EFT. Similarly to the N$^3$LO versus N$^2$LO 3N
forces, these contributions are enhanced by the large $c_i$ values,
and Kaiser found for these partial N$^4$LO 4N contributions a larger
energy of $\sim 2 \mev$ per particle at saturation density.

Finally, we compare our results for symmetric matter with first
calculations of the 4N contributions to the $^4$He ground-state
energy. These were studied in Ref.~\cite{alpha2006} perturbatively
based on the same N$^3$LO 4N forces. We agree with the sign of the 4N
contributions for all topologies and obtain a similar total energy
correction when taking a density $\sim n_0/3$. Also, the estimate of
Ref.~\cite{Riska} for the $V^e$ 4N contribution to the $^4$He
ground-state energy gave $-18 \kev$ per particle, which is of the same
order as our results at $\sim n_0/3$.

\section{Summary and outlook}

We have presented details and additional results of the first complete
N$^3$LO calculation of the neutron-matter energy based on chiral EFT
NN, 3N, and 4N interactions~\cite{fullN3LO}. Our results for the
energy per particle at saturation density give a range of $14.1 - 21
\mev$, which includes uncertainties from different NN potentials, from
the $c_1$ and $c_3$ couplings in 3N forces (these dominate), from
varying the cutoff in many-body forces, and from the uncertainties in
the perturbative many-body expansion around Hartree Fock. For more
systematic studies, it will be important to develop NN potentials that
explore the different $c_i$ couplings.

We have found large contributions to the energy from N$^3$LO 3N forces
in topologies where $\Delta$ excitations are important. Therefore, an
improved EFT convergence is expected in chiral EFT with explicit
$\Delta$ degrees of freedom. In contrast, contributions from the
leading 4N forces are found to be small (see also
Refs.~\cite{Fiorilla2011,Kaiser}). We have presented a first
estimate for the N$^3$LO many-body contributions to the energy of
symmetric nuclear matter, where also large N$^3$LO 3N forces and small
leading 4N forces are found. Our results for symmetric matter show
that the inclusion of N$^3$LO 3N forces will be important in nuclear
structure calculations, and that it is crucial to go beyond the
Hartree-Fock approximation.

Recently, first Quantum Monte Carlo calculations with chiral EFT
interactions are providing nonperturbative benchmarks for neutron
matter and validate the perturbative expansion for chiral NN
potentials with low cutoffs~\cite{QMC_chiral}. Extending these
calculations to 3N forces and N$^3$LO will be important. In addition, the
many-body uncertainties can be reduced in the future by a similarity
RG evolution of NN and 3N forces~\cite{3Nevolution,nm_3Nevolution},
which improves the many-body convergence and will also enable studies
with the chiral NN interactions, which were found to be
nonperturbative in the present calculations.

In addition, we have discussed the impact of our results for
astrophysics: The predicted ranges for the symmetry energy $S_v$ and
its density derivative $L$ are $S_v=28.9-34.9\, \mev$ and $L=43.0-66.6
\, \mev$, which are consistent with recent experimental
constraints~\cite{LL,Kortelainen2010,Tamii2011}. Many of the equations of
state for core-collapse supernova simulations were found to be
inconsistent with the N$^3$LO neutron-matter band.  By extending our
neutron-matter results to neutron-star matter and to high densities,
we confirm the predicted radius range of $9.7-13.9 \km$ for a $1.4 \,
M_{\odot}$ neutron star~\cite{nstar_long} and find a maximal neutron
star mass of $3.1 \, M_{\odot}$.

\begin{acknowledgments}

We thank E.~Epelbaum, R.~J.~Furnstahl, M.~Hempel, N.~Kaiser,
A.~Kleiner, H.~Krebs, J.~M.~Lattimer, C.~J.~Pethick, and G.~Shen for
discussions. This work was supported by the Helmholtz Alliance Program
of the Helmholtz Association, contract HA216/EMMI ``Extremes of
Density and Temperature: Cosmic Matter in the Laboratory'', by the ERC
Grant No.~307986 STRONGINT, the DFG through Grant SFB 634, and by NSF
Grant No.~PHY--1002478.

\end{acknowledgments}

\begin{widetext}

\appendix
\section{N$^\text{3}$LO neutron-matter matrix elements}
\label{app:nm}

In this appendix we present the 3N and 4N matrix elements defined as
\begin{equation}
\langle V_{A\text{N}}\rangle = \frac{1}{A!}\sum_{\sigma_1,\ldots,\sigma_A}\bra{1\cdots A}\mathcal{A}_A\sum_{i_1 \neq \ldots \neq i_A}V_{A\text{N}}(i_1,\ldots,i_A)\ket{1\cdots A}\,,
\end{equation}
entering the neutron-matter Hartree-Fock calculation [see
Eq.~(\ref{eq:hf})] of the N$^3$LO many-body forces.

We use the short-hand notation for the momentum transfer
$\mathbf{k}_{ij} = \mathbf{k}_i - \mathbf{k}_j$,
$\mathbf{k}_{(ij)(kl)} = \mathbf{k}_{ij} + \mathbf{k}_{kl}$ and
$\mathbf{P}_{ij} = \frac{\mathbf{k}_i + \mathbf{k}_j}{2}$, and pion
propagators $K_{ij} = k_{ij}^2 + m_\pi^2$ and $K_{(ij)(kl)} =
k_{(ij)(kl)}^2 + m_\pi^2$.

\subsection{Two-pion-exchange 3N}

\begin{align}
\langle V_{\text{3N}}^{2 \pi} \rangle &= \frac{g_A^2}{f_{\pi}^4} \left( -2\delta c_1 m_{\pi}^2 \left[\frac{\mathbf{k}_{12}\cdot \mathbf{k}_{23}}{K_{12}K_{23}}+\frac{k_{12}^2}{K_{12}^2} \right] +\delta c_3\left[\frac{(\mathbf{k}_{12}\cdot \mathbf{k}_{23})^2}{K_{12}K_{23}}-\frac{k_{12}^4}{K_{12}^2} \right] \right) \nonumber \\
&\quad +k_{13}^2 \, F_{2\pi,1}^{(4)}(-\mathbf{k}_{13},\mathbf{k}_{13})-\mathbf{k}_{12}\cdot \mathbf{k}_{13} \, F_{2\pi,1}^{(4)}(-\mathbf{k}_{12},\mathbf{k}_{13})\,,
\end{align}
with shifts in the low-energy couplings $\delta c_1 = -0.13\gevi$ and
$\delta c_3 = 0.89\gevi$ (see Ref.~\cite{N3LOlong}) and the function
\begin{align}
F_{2\pi,1}^{(4)}(\mathbf{q}_1,\mathbf{q}_2) &= \frac{3 g_A^4}{32 \pi f_{\pi}^6(q_1^2+m_\pi^2)(q_2^2+m_\pi^2)}\Bigl[ m_{\pi}(m_{\pi}^2+3q_1^2+3q_2^2+4\mathbf{q}_1 \cdot \mathbf{q}_2) \nonumber \\
&\quad + (2m_{\pi}^2+q_1^2+q_2^2+2\mathbf{q}_1 \cdot \mathbf{q}_2)(3m_{\pi}^2+3q_1^2+3q_2^2+4\mathbf{q}_1 \cdot \mathbf{q}_2)A(|\mathbf{q}_1+\mathbf{q}_2|) \Bigr] \,,
\end{align}
where $A(q) = 1/(2q)\arctan[q/(2m_{\pi})]$ denotes the loop 
function~\cite{N3LOlong}.

\subsection{Two-pion--one-pion-exchange 3N}

\begin{align}
\langle V_{\text{3N}}^{2\pi 1 \pi} \rangle &= 4 \left[F_1(k_{12})\frac{(\mathbf{k}_{12}\cdot \mathbf{k}_{13})^2}{K_{13}} - F_2(k_{12})\frac{\mathbf{k}_{12}\cdot \mathbf{k}_{13}}{K_{13}} - F_3(0)\frac{k_{23}^2}{K_{23}} + F_3(k_{12})\frac{k_{13}^2}{K_{13}} - F_7(0) \frac{k_{23}^2}{K_{23}} \nonumber \right. \\
&\quad \left. + F_4(k_{12})\frac{(\mathbf{k}_{12}\cdot \mathbf{k}_{13})^2}{K_{13}} + F_5(k_{12}) \frac{k_{13}^2}{K_{13}} -  F_6(k_{12}) \frac{\mathbf{k}_{12}\cdot \mathbf{k}_{13}}{K_{13}} + F_7(k_{12})\frac{k_{13}^2}{K_{13}}\right] \,,
\end{align}
with structure functions $F_1(q)$ to $F_7(q)$ defined in Eqs.~(2.17)--(2.20) of Ref.~\cite{N3LOlong}.

\subsection{Pion-ring 3N}

\begin{align}
\langle V_{\text{3N}}^{\textrm{ring}} \rangle &= 4 \Bigl[-3\, R_1(\mathbf{k}_{12},0)+3\, R_1(\mathbf{k}_{12},\mathbf{k}_{23}) - k_{12}^2\, R_2(\mathbf{k}_{12},0) + k_{12}^2 \, R_2(\mathbf{k}_{12},\mathbf{k}_{23}) +  \mathbf{k}_{12}\cdot \mathbf{k}_{23} \, R_3(\mathbf{k}_{12},\mathbf{k}_{23}) \nonumber \\
&\quad + \mathbf{k}_{12}\cdot \mathbf{k}_{23} \, R_4(\mathbf{k}_{12},\mathbf{k}_{23}) + k_{23}^2\, R_5(\mathbf{k}_{12},\mathbf{k}_{23}) + 2 R_6(0,0) - R_6(\mathbf{k}_{12},0) - R_6(0,\mathbf{k}_{12}) - R_6(-\mathbf{k}_{12},\mathbf{k}_{12})\nonumber \\
&\quad + R_6(\mathbf{k}_{12},\mathbf{k}_{23}) - k_{12}^2\, R_7(-\mathbf{k}_{12},\mathbf{k}_{12}) + k_{12}^2 \, R_7(\mathbf{k}_{12},\mathbf{k}_{23}) + k_{12}^2\, R_8(-\mathbf{k}_{12},\mathbf{k}_{12}) + \mathbf{k}_{12}\cdot \mathbf{k}_{23}\, R_8(\mathbf{k}_{12},\mathbf{k}_{23})\nonumber \\
&\quad + k_{12}^2\, R_9(-\mathbf{k}_{12},\mathbf{k}_{12}) + \mathbf{k}_{12}\cdot \mathbf{k}_{23} \, R_9(\mathbf{k}_{12},\mathbf{k}_{23}) - 3 R_{10}(-\mathbf{k}_{12},\mathbf{k}_{12}) + 3 R_{10}(\mathbf{k}_{12},\mathbf{k}_{23})
+ 2 S_1(0,0)\nonumber \\
&\quad - S_1(\mathbf{k}_{12},0) - S_1(0,\mathbf{k}_{12}) - S_1(-\mathbf{k}_{12},\mathbf{k}_{12}) + S_1(\mathbf{k}_{12},\mathbf{k}_{23}) - k_{12}^2 S_2(-\mathbf{k}_{12},\mathbf{k}_{12}) + k_{12}^2 S_2(\mathbf{k}_{12},\mathbf{k}_{23})\nonumber \\
&\quad + k_{12}^2 S_3(-\mathbf{k}_{12},\mathbf{k}_{12}) + \mathbf{k}_{12}\cdot \mathbf{k}_{23} S_3(\mathbf{k}_{12},\mathbf{k}_{23}) + k_{12}^2 S_4(-\mathbf{k}_{12},\mathbf{k}_{12}) + \mathbf{k}_{12}\cdot \mathbf{k}_{23} S_4(\mathbf{k}_{12},\mathbf{k}_{23})\nonumber \\
&\quad - k_{12}^2 S_5(-\mathbf{k}_{12},\mathbf{k}_{12}) + k_{23}^2 S_5(\mathbf{k}_{12}, \mathbf{k}_{23}) - 3 S_6(-\mathbf{k}_{12},\mathbf{k}_{12}) + 3 S_6(\mathbf{k}_{12},\mathbf{k}_{23}) \Bigr] \,,
\end{align}
where the structure functions $R_i$ and $S_i$ are defined in Eqs.~(A2)
and (A7) of Ref.~\cite{N3LOlong}.

\subsection{Two-pion-exchange--contact 3N}

\begin{equation}
\langle V_{\text{3N}}^{2\pi\text{-cont}}\rangle = -\frac{g_A^2}{2\pi f_{\pi}^4} \, C_T \left(g_A^2 \left[\frac{3m_{\pi}}{4}+\frac{m_{\pi}^3}{4m_{\pi}^2+k_{12}^2}-2(2m_{\pi}^2+k_{12}^2)A(k_{12})\right] - \left[\frac{m_\pi}{2} - (2m_\pi^2+k_{12}^2)A(k_{12})\right]\right) \,.
\end{equation}

\subsection{Relativistic-corrections 3N}

\begin{align}
\langle V_{\text{3N}}^{1/m}\rangle &= 2 \left[k_{12}^2 F_{1/m}^1(\mathbf{k}_{12},\mathbf{k}_{12}) + \mathbf{k}_{12}\cdot \mathbf{k}_{23} F_{1/m}^1(\mathbf{k}_{12},\mathbf{k}_{23}) - (\mathbf{k}_{12}\times \mathbf{k}_{23})^2 F_{1/m}^2(-\mathbf{k}_{12},\mathbf{k}_{13},\mathbf{P}_{12},\mathbf{P}_{23}) \right. \nonumber \\
&\quad + k_{12}^2 F_{1/m}^3(\mathbf{k}_{12},\mathbf{k}_{12})  + \mathbf{k}_{12}\cdot \mathbf{k}_{23} F_{1/m}^3(\mathbf{k}_{12},\mathbf{k}_{23}) - (\mathbf{k}_{12}\times \mathbf{k}_{13})\cdot(\mathbf{k}_{12}\times \mathbf{P}_{23}) F_{1/m}^4(\mathbf{k}_{12},\mathbf{k}_{13}) \nonumber\\
&\quad - (\mathbf{k}_{12}\times \mathbf{k}_{13})\cdot (\mathbf{k}_{12}\times \mathbf{P}_{13})F_{1/m}^5(\mathbf{k}_{12},\mathbf{k}_{13}) + k_{12}^2 F_{1/m}^6(\mathbf{k}_{12},\mathbf{k}_{23}) - k_{12}^2 F_{1/m}^7(\mathbf{k}_{12},-\mathbf{k}_{12})\nonumber \\
&\quad \left. + k_{12}^2 F_{1/m}^7(\mathbf{k}_{12},\mathbf{k}_{23}) - k_{12}^2 F_{1/m}^8(\mathbf{k}_{12},\mathbf{P}_{12},\mathbf{P}_{23})- k_{12}^2 F_{1/m}^9(k_{12}) + k_{12}^2 F_{1/m}^{10}(k_{12}) + k_{12}^2 F_{1/m}^{11}(k_{12}) \right] \,,
\end{align}
with
\begin{align}
&F_{1/m}^1(\mathbf{q}_1,\mathbf{q}_2)= -\frac{g_A^4}{16 m f_{\pi}^4}\frac{(1-2\bar{\beta}_8)(\mathbf{q}_1 \cdot \mathbf{q}_2)^2}{(q_1^2+m_{\pi}^2)^2(q_2^2+m_{\pi}^2)}\,,\\[1mm]
&F_{1/m}^2(\mathbf{q}_1,\mathbf{q}_2,\mathbf{q}_3,\mathbf{q}_4)= \frac{g_A^4}{8 m f_{\pi}^4} \frac{(1-2\bar{\beta}_8)\mathbf{q}_1\cdot \mathbf{q}_4+(1+2\bar{\beta}_8)\mathbf{q}_1\cdot \mathbf{q}_3}{(q_1^2+m_{\pi}^2)^2(q_2^2+m_{\pi}^2)}
\,,\\[1mm]
&F_{1/m}^3(\mathbf{q}_1,\mathbf{q}_2)= -\frac{g_A^4}{16 m f_{\pi}^4}\frac{(2\bar{\beta}_9-1)q_1^2}{(q_1^2+m_{\pi}^2)(q_2^2+m_{\pi}^2)}=-F_{1/m}^4(\mathbf{q}_1,\mathbf{q}_2)\frac{q_1^2}{2}=-F_{1/m}^5(\mathbf{q}_1\,,\mathbf{q}_2)\frac{q_1^2(2\bar{\beta}_9-1)}{2(2\bar{\beta}_9+1)}\,,\\[1mm]
&F_{1/m}^6(\mathbf{q}_1,\mathbf{q}_2)= \frac{g_A^2}{4 m f_{\pi}^2}C_S\frac{(1-2\bar{\beta}_8)\mathbf{q}_1\cdot \mathbf{q}_2}{(q_1^2+m_{\pi}^2)^2}=F_{1/m}^7(\mathbf{q}_1,\mathbf{q}_2)\frac{C_S}{C_T}\,,\\[1mm]
&F_{1/m}^8(\mathbf{q}_1,\mathbf{q}_2,\mathbf{q}_3)= \frac{g_A^2}{m f_{\pi}^2}C_T\frac{(1-2\bar{\beta}_8)\mathbf{q}_1\cdot \mathbf{q}_3+(1+2\bar{\beta}_8)\mathbf{q}_1 \cdot \mathbf{q}_2}{(q_1^2+m_{\pi}^2)^2}\,,\\[1mm]
&F_{1/m}^9(q)\,= \frac{g_A^2}{8 m f_{\pi}^2}C_S\frac{2\bar{\beta}_9-1}{q^2+m_{\pi}^2}=F_{1/m}^{10}(q)\frac{C_S}{C_T}=F_{1/m}^{11}(q)\frac{C_S}{2C_T}\,.
\end{align}

\subsection{Three-pion-exchange and pion-interaction 4N}

\begin{align}
\langle V_{\text{4N}}^{a}\rangle &= -\frac{g_A^6}{8 f_\pi^6}\Biggl(\Bigl[(\mathbf{k}_1\times \mathbf{k}_2)\cdot\mathbf{k}_{34} + (\mathbf{k}_3 \times \mathbf{k}_4)\cdot\mathbf{k}_{12}\bigr]^2\left[\frac{1}{K_{14} K_{(14)(23)}^2K_{24}} + \frac{1}{K_{12}K_{14}^2K_{34}} - \frac{1}{K_{12}K_{13}^2K_{14}}\right] \nonumber\\
&\quad + \frac{k_{14}^2(\mathbf{k}_{14}\times\mathbf{k}_{(14)(23)})^2}{K_{14}^2K_{(14)(23)}^2} - \frac{\mathbf{k}_{14}\cdot\mathbf{k}_{24}(\mathbf{k}_{(14)(23)}\times\mathbf{k}_{14})\cdot(\mathbf{k}_{(14)(23)}\times\mathbf{k}_{24})}{K_{14}K_{(14)(23)}^2K_{24}} \nonumber\\
&\quad + \frac{\mathbf{k}_{12}\cdot\mathbf{k}_{34}(\mathbf{k}_{14}\times\mathbf{k}_{12})\cdot(\mathbf{k}_{14}\times\mathbf{k}_{34})}{K_{12}K_{14}^2K_{34}} - \frac{\mathbf{k}_{12}\cdot\mathbf{k}_{14}(\mathbf{k}_{13}\times\mathbf{k}_{12})\cdot(\mathbf{k}_{13}\times\mathbf{k}_{14})}{K_{12}K_{13}^2K_{14}} \vphantom{\int} \Biggr)\,,\\[1mm]
\langle V_{\text{4N}}^{e}\rangle &= \frac{g_A^4}{16 f_\pi^6} \left[ \vphantom{\int } -2\frac{k_{24}^2}{K_{13}K_{24}^2} \mathbf{k}_{13}\cdot(\mathbf{k}_{13}+\mathbf{k}_{24}) \right. - \frac{\mathbf{k}_{13}\cdot\mathbf{k}_{24}}{K_{13}K_{23}K_{24}} \mathbf{k}_{23}\cdot(\mathbf{k}_{13}+\mathbf{k}_{24}) + 2\frac{\mathbf{k}_{13}\cdot\mathbf{k}_{34}}{K_{13}K_{24}K_{34}} \mathbf{k}_{24}\cdot\mathbf{k}_{14}\nonumber\\
&\quad + 2\frac{\mathbf{k}_{23}\cdot\mathbf{k}_{24}}{K_{13}K_{23}K_{24}} \mathbf{k}_{13}\cdot(\mathbf{k}_{13}+\mathbf{k}_{24}) +2\frac{\mathbf{k}_{23}\cdot\mathbf{k}_{14}}{K_{14}K_{23}K_{34}} \mathbf{k}_{34}\cdot\mathbf{k}_{13} +\left.2\frac{\mathbf{k}_{12}\cdot\mathbf{k}_{24}}{K_{12}K_{24}K_{34}} \mathbf{k}_{34}\cdot\mathbf{k}_{23} \vphantom{\int}\right]\,,\\[1mm]
\langle V_{\text{4N}}^{f}\rangle &= \frac{g_A^4}{32f_\pi^6} \left[ \vphantom{\int}(m_\pi^2 + 2K_{(12)(34)})\frac{k_{12}^2 k_{34}^2}{K_{12}^2K_{34}^2}\right. -(K_{(14)(32)} + 2K_{13})\frac{\mathbf{k}_{12}\cdot\mathbf{k}_{34}\,\mathbf{k}_{14}\cdot\mathbf{k}_{23}}{K_{12}K_{14}K_{23}K_{34}}\nonumber\\
&\quad \left.-2(K_{14} + K_{(34)(21)} + K_{23}) \frac{\mathbf{k}_{12}\cdot\mathbf{k}_{24}\,\mathbf{k}_{13}\cdot\mathbf{k}_{34}}{K_{12}K_{13}K_{24}K_{34}}\vphantom{\int}\right] \,.
\end{align}

\section{N$^3$LO symmetric nuclear-matter matrix elements}
\label{app:sym}

We now turn to the 3N and 4N matrix elements defined as
\begin{equation}
\langle V_{A\text{N}}\rangle = \frac{1}{A!}\sum_{\tau_1,\ldots,\tau_A}\sum_{\sigma_1,\ldots,\sigma_A}\bra{1\cdots A}\mathcal{A}_A\sum_{i_1 \neq \ldots \neq i_A}V_{A\text{N}}(i_1,\ldots,i_A)\ket{1\cdots A}\,,
\label{eq:snm_me}
\end{equation}
entering the symmetric nuclear-matter Hartree-Fock calculation of the
N$^3$LO many-body forces.

\subsection{Two-pion-exchange 3N}

\begin{align}
\langle V_{\text{3N}}^{2 \pi} \rangle &= 6 \frac{g_A^2}{f_{\pi}^2} \biggl( -2 \frac{\delta c_1 m_{\pi}^2}{f_{\pi}^2} \left[\frac{\mathbf{k}_{12}\cdot \mathbf{k}_{23}}{K_{12}K_{23}} + 2 \frac{k_{12}^2}{K_{12}^2} \right] +\frac{\delta c_3}{f_{\pi}^2}\left[\frac{(\mathbf{k}_{12}\cdot \mathbf{k}_{23})^2}{K_{12}K_{23}} - 2 \frac{k_{12}^4}{K_{12}^2} \right] - \frac{\delta c_4}{f_{\pi}^2} \, \frac{(\mathbf{k}_{12} \times \mathbf{k}_{23})^2}{K_{12}K_{23}} \biggr) \nonumber \\
&\quad + 6 \left[2\mathbf{k}_{13}^2  F_{2\pi ,1}^{(4)}(-\mathbf{k}_{13},\mathbf{k}_{13}) - \mathbf{k}_{12} \cdot \mathbf{k}_{13}  F_{2\pi ,1}^{(4)}(-\mathbf{k}_{12},\mathbf{k}_{13})\right] - (\mathbf{k}_{12} \times \mathbf{k}_{13})^2 F_{2\pi ,2}^{(4)}(-\mathbf{k}_{12},\mathbf{k}_{13})\,,
\end{align}
with shifts in the low-energy couplings $\delta c_1$, $\delta c_3 =
-\delta c_4$, the function $F_{2\pi,1}^{(4)}$ is as given in
Appendix~\ref{app:nm}, and
\begin{equation}
F_{2\pi,2}^{(4)}(\mathbf{q}_1,\mathbf{q}_2) = -\frac{9 g_A^4}{8 \pi f_{\pi}^6(q_1^2+m_\pi^2)(q_2^2+m_\pi^2)}\Bigl[ m_{\pi}+ (4m_{\pi}^2+q_1^2+q_2^2+2\mathbf{q}_1 \cdot \mathbf{q}_2)A(|\mathbf{q}_1+\mathbf{q}_2|)\Bigr] \,.
\end{equation}

\subsection{Two-pion--one-pion-exchange 3N}

\begin{align}
\langle V_{\text{3N}}^{2\pi 1 \pi} \rangle &= 24 \biggl[F_1(k_{12})\frac{(\mathbf{k}_{12}\cdot \mathbf{k}_{13})^2}{K_{13}} - F_2(k_{12})\frac{\mathbf{k}_{12}\cdot \mathbf{k}_{13}}{K_{13}} + F_3(k_{12})\frac{k_{13}^2}{K_{13}} + F_4(k_{12})\frac{(\mathbf{k}_{12}\cdot \mathbf{k}_{13})^2}{K_{13}} \nonumber \\
&\quad + F_5(k_{12}) \frac{k_{13}^2}{K_{13}} - F_6(k_{12}) \frac{\mathbf{k}_{12}\cdot \mathbf{k}_{13}}{K_{13}} - 2 F_7(0) \frac{k_{23}^2}{K_{23}} + F_7(k_{12})\frac{k_{13}^2}{K_{13}} + 4 F_8(k_{12})\frac{\mathbf{k}_{12}\cdot \mathbf{k}_{13}}{K_{13}}\biggr] \,,
\end{align}
with structure functions $F_1(q)$ to $F_8(q)$ defined in
Eqs.~(2.17)--(2.20) of Ref.~\cite{N3LOlong}.

\subsection{Pion-ring 3N}

\begin{align}
\langle V_{\text{3N}}^{\textrm{ring}}\rangle &= 8 \Bigl[9 R_1(-\mathbf{k}_{12},\mathbf{k}_{13}) + 3 k_{12}^2 \, R_2(-\mathbf{k}_{12},\mathbf{k}_{13}) - 3 \mathbf{k}_{12}\cdot \mathbf{k}_{13} \, R_3(-\mathbf{k}_{12},\mathbf{k}_{13}) -3 \mathbf{k}_{12}\cdot \mathbf{k}_{13} \, R_4(-\mathbf{k}_{12},\mathbf{k}_{13})\nonumber \\
&\quad + 3 k_{13}^2\, R_5(-\mathbf{k}_{12},\mathbf{k}_{13}) - 6 R_6(-\mathbf{k}_{13},\mathbf{k}_{13}) + 3 R_6(-\mathbf{k}_{12},\mathbf{k}_{13}) - 2 k_{13}^2\, R_7(-\mathbf{k}_{13},\mathbf{k}_{13}) + k_{12}^2 \, R_7(-\mathbf{k}_{12},\mathbf{k}_{13})\nonumber \\
&\quad + 2 k_{13}^2\, R_8(-\mathbf{k}_{13},\mathbf{k}_{13}) - \mathbf{k}_{12}\cdot \mathbf{k}_{13}\, R_8(-\mathbf{k}_{12},\mathbf{k}_{13}) + 2 k_{13}^2\, R_9(-\mathbf{k}_{13},\mathbf{k}_{13}) - \mathbf{k}_{12}\cdot \mathbf{k}_{13} \, R_9(-\mathbf{k}_{12},\mathbf{k}_{13}) \nonumber \\
&\quad - 6 R_{10}(-\mathbf{k}_{13},\mathbf{k}_{13}) + 3 R_{10}(-\mathbf{k}_{12},\mathbf{k}_{13}) - 6 S_1(-\mathbf{k}_{12},0) + 3 S_1(-\mathbf{k}_{12},\mathbf{k}_{13}) + 3 k_{12}^2 S_2(-\mathbf{k}_{12},\mathbf{k}_{13})\nonumber \\
&\quad - 3 \mathbf{k}_{12}\cdot \mathbf{k}_{13} S_3(-\mathbf{k}_{12},\mathbf{k}_{13}) - 3 \mathbf{k}_{12}\cdot \mathbf{k}_{13} S_4(-\mathbf{k}_{12},\mathbf{k}_{13}) + 3 k_{13}^2 S_5(-\mathbf{k}_{12}, \mathbf{k}_{13}) + 9 S_6(-\mathbf{k}_{12},\mathbf{k}_{13}) \Bigr] \,,
\end{align}
with structure functions $R_i$ and $S_i$ defined in Eqs.~(A2) and (A7)
of Ref.~\cite{N3LOlong}.

\subsection{Two-pion-exchange--contact 3N}

\begin{equation}
\langle V_{\text{3N}}^{2\pi\text{-cont}}\rangle = \frac{3 g_A^2}{\pi f_{\pi}^4} \, C_T \left( g_A^2 \biggl[3m_{\pi}-\frac{m_{\pi}^3}{3m_{\pi}^2+K_{12}}+(4m_{\pi}^2-3k_{12}^2)A(k_{12}) \biggr] - \Bigl[m_{\pi}+(2m_{\pi}^2+k_{12}^2)A(k_{12})\Bigr] \right)\,.
\end{equation}

\subsection{Relativistic-corrections 3N}

\begin{align}
\langle V_{\text{3N}}^{1/m}\rangle &= 12 \left[2 k_{13}^2 F_{1/m}^1(-\mathbf{k}_{13},\mathbf{k}_{13}) - \mathbf{k}_{12}\cdot \mathbf{k}_{13} F_{1/m}^1(-\mathbf{k}_{12},\mathbf{k}_{13}) - (\mathbf{k}_{12}\times \mathbf{k}_{13})^2 F_{1/m}^2(-\mathbf{k}_{12},\mathbf{k}_{13},\mathbf{P}_{12},\mathbf{P}_{23}) \right. \nonumber \\
&\quad - (\mathbf{k}_{12}\times \mathbf{k}_{13})\cdot (\mathbf{k}_{12}\times \mathbf{P}_{23}) F_{1/m}^3(-\mathbf{k}_{12},\mathbf{k}_{13}) - (\mathbf{k}_{12}\times \mathbf{P}_{13})\cdot (\mathbf{k}_{12}\times \mathbf{k}_{13}) F_{1/m}^4(-\mathbf{k}_{12},\mathbf{k}_{13})\nonumber \\
&\quad - \mathbf{k}_{12}\cdot \mathbf{k}_{13} F_{1/m}^5(-\mathbf{k}_{12},\mathbf{k}_{13},\mathbf{P}_{12},\mathbf{P}_{23},\mathbf{P}_{13}) + (\mathbf{k}_{12}\times \mathbf{k}_{13})^2 F_{1/m}^6(-\mathbf{k}_{12},\mathbf{k}_{13}) - \mathbf{k}_{12}\cdot \mathbf{P}_{13} F_{1/m}^7(-\mathbf{k}_{12},\mathbf{k}_{13})\nonumber \\
&\quad + k_{12}^2 F_{1/m}^8(-\mathbf{k}_{12},\mathbf{k}_{13}) + k_{12}^2 F_{1/m}^9(-\mathbf{k}_{12},\mathbf{k}_{13}) + k_{12}^2 F_{1/m}^{10}(-\mathbf{k}_{12},\mathbf{P}_{12},\mathbf{P}_{23}) - \mathbf{k}_{12}\cdot \mathbf{k}_{13} F_{1/m}^{11}(k_{12})\nonumber \\
&\quad \left. - \mathbf{k}_{12}\cdot \mathbf{k}_{13} F_{1/m}^{12}(k_{12}) - \mathbf{k}_{12}\cdot \mathbf{P}_{23} F_{1/m}^{13}(k_{12}) - \mathbf{k}_{12}\cdot \mathbf{P}_{12} F_{1/m}^{14}(k_{12}) \right]\,,
\end{align}
with
\begin{align}
&F_{1/m}^1(\mathbf{q}_1,\mathbf{q}_2)= -\frac{g_A^4}{16 m f_{\pi}^4} \frac{1}{(q_1^2+m_{\pi}^2)(q_2^2+m_{\pi}^2)} \left[ \frac{1}{(q_1^2+m_{\pi}^2)}(1-2\bar{\beta}_8) (\mathbf{q}_1\cdot \mathbf{q}_2)^2 + (2\bar{\beta}_9-1) q_1^2 \right]\,,\\[1mm]
&F_{1/m}^2(\mathbf{q}_1,\mathbf{q}_2,\mathbf{q}_3,\mathbf{q}_4)= \frac{g_A^2}{8 m f_{\pi}^4} \frac{1}{(q_1^2+m_{\pi}^2)(q_2^2+m_{\pi}^2)} \left(\frac{g_A^2}{(q_1^2+m_{\pi}^2)}\Bigl[(1-2\bar{\beta}_8) \mathbf{q}_1\cdot \mathbf{q}_4 +(1+2\bar{\beta}_8) \mathbf{q}_1\cdot \mathbf{q}_3 \Bigr]\right)\,,\\[1mm]
&F_{1/m}^3(\mathbf{q}_1,\mathbf{q}_2)= -\frac{g_A^4}{8 m f_{\pi}^4} \frac{2\bar{\beta}_9-1}{(q_1^2+m_{\pi}^2)(q_2^2+m_{\pi}^2)}=-F_{1/m}^4(\mathbf{q}_1,\mathbf{q}_2)\frac{2\bar{\beta}_9-1}{2\bar{\beta}_9+1}\,,\\[1mm]
&F_{1/m}^5(\mathbf{q}_1,\mathbf{q}_2,\mathbf{q}_3,\mathbf{q}_4,\mathbf{q}_5)= \frac{g_A^2}{4 m f_{\pi}^4} \frac{1}{(q_1^2+m_{\pi}^2)(q_2^2+m_{\pi}^2)} \biggl(-\frac{g_A^2}{(q_1^2+m_{\pi}^2)}\mathbf{q}_1\cdot \mathbf{q}_2 \Bigl[(1-2\bar{\beta}_8)\mathbf{q}_1\cdot \mathbf{q}_4+(1+2\bar{\beta}_8)\mathbf{q}_1\cdot \mathbf{q}_3 \Bigr]\nonumber\\
&\quad\quad\quad\quad\quad\quad\quad\quad\quad\quad\quad + \mathbf{q}_2\cdot (\mathbf{q}_5-\mathbf{q}_4) +g_A^2(2\bar{\beta}_9-1) \mathbf{q}_1\cdot \mathbf{q}_4 \biggr)\,, \\[1mm]
&F_{1/m}^6(\mathbf{q}_1,\mathbf{q}_2)= -\frac{g_A^2}{8 m f_{\pi}^4}\frac{1}{(q_1^2+m_{\pi}^2)(q_2^2+m_{\pi}^2)}\left[\frac{g_A^2}{q_1^2+m_{\pi}^2}(1-2\bar{\beta}_8)\, \mathbf{q}_1\cdot \mathbf{q}_2 + 1 \right]\,,\\[1mm]
&F_{1/m}^7(\mathbf{q}_1,\mathbf{q}_2)= -\frac{g_A^4}{4 m f_{\pi}^4}\frac{(2\bar{\beta}_9+1)\, \mathbf{q}_1\cdot \mathbf{q}_2}{(q_1^2+m_{\pi}^2)(q_2^2+m_{\pi}^2)}\,,\\[1mm]
&F_{1/m}^8(\mathbf{q}_1,\mathbf{q}_2)= \frac{g_A^2}{4 m f_{\pi}^2} C_S \frac{(1-2\bar{\beta}_8)\, \mathbf{q}_1\cdot \mathbf{q}_2}{(q_1^2+m_{\pi}^2)^2} =F_{1/m}^9(\mathbf{q}_1,\mathbf{q}_2)\frac{C_S}{C_T}\,,\\[1mm]
&F_{1/m}^{10}(\mathbf{q}_1,\mathbf{q}_2,\mathbf{q}_3)= \frac{g_A^2}{m f_{\pi}^2}\frac{1}{(q_1^2+m_{\pi}^2)^2} C_T \Bigl[ (1-2\bar{\beta}_8)\, \mathbf{q}_1\cdot \mathbf{q}_3+(1+2\bar{\beta}_8)\, \mathbf{q}_1\cdot \mathbf{q}_2 \Bigr]\,,\\
&F_{1/m}^{11}(q)= \frac{g_A^2}{4 m f_{\pi}^2}\frac{2\bar{\beta}_9-1}{q^2+m_{\pi}^2} C_S=F_{1/m}^{12}(q)\frac{C_S}{C_T}=F_{1/m}^{13}(q)\frac{C_S}{4C_T}=-F_{1/m}^{14}(q)\frac{C_S(2\bar{\beta}_9-1)}{4C_T(2\bar{\beta}_9+1)}\,.
\end{align}

\subsection{Three-pion-exchange and pion-interaction 4N}

\begin{align}
\langle V_{\text{4N}}^{a}\rangle &= -\frac{3 g_A^6}{4 f_\pi^6}\Biggl( 4\left[\frac{k_{14}^2(\mathbf{k}_{14}\cdot\mathbf{k}_{(14)(23)})^2}{K_{14}^2K_{(14)(23)}^2}+\frac{\mathbf{k}_{12}\cdot\mathbf{k}_{34}\,\mathbf{k}_{12}\cdot\mathbf{k}_{14}\,\mathbf{k}_{34}\cdot\mathbf{k}_{14}}{K_{12}K_{14}^2K_{34}}-\frac{\mathbf{k}_{12}\cdot\mathbf{k}_{14}\,\mathbf{k}_{12}\cdot\mathbf{k}_{13}\,\mathbf{k}_{14}\cdot\mathbf{k}_{13}}{K_{12}K_{13}^2K_{14}}\right]\nonumber\\
&\quad + 2 \left[\frac{(\mathbf{k}_{14}\times\mathbf{k}_{42})\cdot(\mathbf{k}_{31}\times\mathbf{k}_{42})\,\mathbf{k}_{14}\cdot\mathbf{k}_{(14)(23)}}{K_{14}K_{(14)(23)}^2K_{24}} - \frac{(\mathbf{k}_{12}\times\mathbf{k}_{43})\cdot(\mathbf{k}_{31}\times\mathbf{k}_{43})\,\mathbf{k}_{12}\cdot\mathbf{k}_{14}}{K_{12}K_{14}^2K_{34}} \right.\nonumber\\
&\quad\quad\quad \left.+ \frac{(\mathbf{k}_{12}\times\mathbf{k}_{41})\cdot(\mathbf{k}_{34}\times\mathbf{k}_{41})\,\mathbf{k}_{12}\cdot\mathbf{k}_{13}}{K_{12}K_{13}^2K_{14}}\right] \nonumber\\
&\quad- 2 \left[\frac{(\mathbf{k}_{14}\times\mathbf{k}_{42})\cdot(\mathbf{k}_{14}\times\mathbf{k}_{23})\,\mathbf{k}_{42}\cdot\mathbf{k}_{(14)(23)}}{K_{14}K_{(14)(23)}^2K_{24}} - \frac{(\mathbf{k}_{12}\times\mathbf{k}_{43})\cdot(\mathbf{k}_{12}\times\mathbf{k}_{24})\,\mathbf{k}_{43}\cdot\mathbf{k}_{14}}{K_{12}K_{14}^2K_{34}} \right.\nonumber\\
&\quad\quad\quad \left.+ \frac{(\mathbf{k}_{12}\times\mathbf{k}_{41})\cdot(\mathbf{k}_{12}\times\mathbf{k}_{23})\,\mathbf{k}_{41}\cdot\mathbf{k}_{13}}{K_{12}K_{13}^2K_{14}}\right] \nonumber\\
&\quad + \Bigl[(\mathbf{k}_1\times \mathbf{k}_2)\cdot\mathbf{k}_{34} + (\mathbf{k}_3 \times \mathbf{k}_4)\cdot\mathbf{k}_{12}\Bigr]^2 \left[\frac{1}{K_{14} K_{(14)(23)}^2K_{24}} + \frac{1}{K_{12}K_{14}^2K_{34}} - \frac{1}{K_{12}K_{13}^2K_{14}}\right] \nonumber\\
&\quad + 2 \frac{k_{14}^2(\mathbf{k}_{14}\times\mathbf{k}_{(14)(23)})^2}{K_{14}^2K_{(14)(23)}^2} - \frac{\mathbf{k}_{14}\cdot\mathbf{k}_{24}(\mathbf{k}_{(14)(23)}\times\mathbf{k}_{14})\cdot(\mathbf{k}_{(14)(23)}\times\mathbf{k}_{24})}{K_{14}K_{(14)(23)}^2K_{24}} \nonumber\\
&\quad + \frac{\mathbf{k}_{12}\cdot\mathbf{k}_{34}(\mathbf{k}_{14}\times\mathbf{k}_{12})\cdot(\mathbf{k}_{14}\times\mathbf{k}_{34})}{K_{12}K_{14}^2K_{34}} - \frac{\mathbf{k}_{12}\cdot\mathbf{k}_{14}(\mathbf{k}_{13}\times\mathbf{k}_{12})\cdot(\mathbf{k}_{13}\times\mathbf{k}_{14})}{K_{12}K_{13}^2K_{14}}\Biggr) \,,
\end{align}
\begin{align}
\langle V_{\text{4N}}^{c}\rangle &= \frac{3 g_A^4}{2 f_\pi^6}\Biggl( 2\left[\frac{k_{14}^2\,\mathbf{k}_{14}\cdot\mathbf{k}_{(14)(23)}}{K_{14}^2K_{(14)(23)}}+\frac{\mathbf{k}_{12}\cdot\mathbf{k}_{34}\,\mathbf{k}_{34}\cdot\mathbf{k}_{14}}{K_{12}K_{14}K_{34}}-\frac{\mathbf{k}_{12}\cdot\mathbf{k}_{14}\,\mathbf{k}_{14}\cdot\mathbf{k}_{13}}{K_{12}K_{13}K_{14}}\right] \nonumber\\
&\quad+ \left[\frac{(\mathbf{k}_{14}\times\mathbf{k}_{42})\cdot(\mathbf{k}_{31}\times\mathbf{k}_{42})}{K_{14}K_{(14)(23)}K_{24}} - \frac{(\mathbf{k}_{12}\times\mathbf{k}_{43})\cdot(\mathbf{k}_{31}\times\mathbf{k}_{43})}{K_{12}K_{14}K_{34}} + \frac{(\mathbf{k}_{12}\times\mathbf{k}_{41})\cdot(\mathbf{k}_{34}\times\mathbf{k}_{41})}{K_{12}K_{13}K_{14}}\right]\Biggr)\,,\\[1mm]
\langle V_{\text{4N}}^{e}\rangle &= \frac{3g_A^4}{8 f_\pi^6} \left[ \vphantom{\int } -4\frac{k_{24}^2}{K_{13}K_{24}^2} \mathbf{k}_{13}\cdot(\mathbf{k}_{13}+\mathbf{k}_{24}) \right. + \frac{\mathbf{k}_{13}\cdot\mathbf{k}_{24}}{K_{13}K_{23}K_{24}} \mathbf{k}_{23}\cdot(\mathbf{k}_{13}+\mathbf{k}_{24}) + 6\frac{\mathbf{k}_{13}\cdot\mathbf{k}_{34}}{K_{13}K_{24}K_{34}} \mathbf{k}_{24}\cdot\mathbf{k}_{14}\nonumber\\
&\quad - 2\frac{\mathbf{k}_{23}\cdot\mathbf{k}_{24}}{K_{13}K_{23}K_{24}} \mathbf{k}_{13}\cdot(\mathbf{k}_{13}+\mathbf{k}_{24}) +6\frac{\mathbf{k}_{23}\cdot\mathbf{k}_{14}}{K_{14}K_{23}K_{34}} \mathbf{k}_{34}\cdot\mathbf{k}_{13} +\left.6\frac{\mathbf{k}_{12}\cdot\mathbf{k}_{24}}{K_{12}K_{24}K_{34}} \mathbf{k}_{34}\cdot\mathbf{k}_{23} \vphantom{\int}\right]\,,\\[1mm]
\langle V_{\text{4N}}^{f}\rangle &= \frac{3g_A^4}{16f_\pi^6} \left[ \vphantom{\int} (6m_\pi^2 + 4K_{(12)(34)})\frac{k_{12}^2 k_{34}^2}{K_{12}^2K_{34}^2}\right. +(K_{(14)(32)} - 6K_{13})\frac{\mathbf{k}_{12}\cdot\mathbf{k}_{34}\,\mathbf{k}_{14}\cdot\mathbf{k}_{23}}{K_{12}K_{14}K_{23}K_{34}}\nonumber\\
&\quad \left.-(6K_{14} -2K_{(34)(21)} + 6K_{23}) \frac{\mathbf{k}_{12}\cdot\mathbf{k}_{24}\,\mathbf{k}_{13}\cdot\mathbf{k}_{34}}{K_{12}K_{13}K_{24}K_{34}}\vphantom{\int}\right] \,.
\end{align}

\subsection{Two-pion-exchange--contact 4N}

\begin{align}
\langle V_{\text{4N}}^{k}\rangle &= - C_T \, \frac{12g_A^4}{f_\pi^4}\left[\frac{k_{13}^2k_{24}^2-(\mathbf{k}_{13}\cdot\mathbf{k}_{24})^2}{K_{13}K_{(13)(24)}^2}-2\frac{(\mathbf{k}_{13}\cdot\mathbf{k}_{(13)(24)})^2}{K_{13}K_{(13)(24)}^2}\right]\,,\\[1mm]
\langle V_{\text{4N}}^{l}\rangle &= -C_T \, \frac{12g_A^2}{f_\pi^4} \frac{\mathbf{k}_{13}\cdot\mathbf{k}_{(13)(24)}}{K_{13}K_{(13)(24)}}\,,\\[1mm]
\langle V_{\text{4N}}^{n}\rangle &= - C_T^2 \, \frac{12g_A^2}{f_\pi^4} \frac{\mathbf{k}_{(13)(24)}^2}{K_{(13)(24)}^2}\,.
\end{align}
\end{widetext}

\end{document}